\documentclass[pre,twocolumn,showpacs,amsmath,amssymb,floatfix]{revtex4}

\usepackage{graphicx}% Include figure files
\usepackage{dcolumn} % Align table columns on decimal point
\usepackage{bm}      % Bold math
\usepackage{color}
\definecolor{orange}{cmyk}{0,0.6,0.8,0}
\begin{document}

\title{Structures in magnetohydrodynamic turbulence: detection and scaling}
\author{V. M. Uritsky$^1$, A. Pouquet$^2$, D. Rosenberg$^2$, P.D. Mininni$^{2,3}$, and E. Donovan $^1$}
\affiliation{$^1$ Physics and Astronomy Department, University of Calgary, Calgary, AB T2N1N4 Canada \\
             $^2$ NCAR, P.O. Box 3000, Boulder, Colorado 80307-3000, U.S.A.\\
             $^3$ Departamento de F\'\i sica, Facultad de Ciencias Exactas y Naturales, Universidad de Buenos Aires and IFIBA, CONICET, Ciudad Universitaria, 1428 Buenos Aires, Argentina. 
            }
\date{\today}

\begin{abstract}
We present a systematic analysis of statistical properties of turbulent current and vorticity structures at a given time using cluster analysis. The data stems from numerical simulations of decaying three-dimensional (3D) magnetohydrodynamic turbulence in the absence of an imposed uniform magnetic field; the magnetic Prandtl number is taken equal to unity, and we use a periodic box with grids of up to $1536^3$ points, and with Taylor Reynolds numbers up to $1100$. The initial conditions are either an X-point  configuration embedded in 3D, the so-called Orszag-Tang vortex, or an Arn'old-Beltrami-Childress configuration with a fully helical velocity and magnetic field. In each case two snapshots are analyzed, separated by one turn-over time, starting just after the peak of dissipation. We show that the algorithm is able to select a large number of structures (in excess of $8,000$) for each snapshot and that the statistical properties of these clusters are remarkably similar for the two snapshots as well as for the two flows under study in terms of scaling laws for the cluster characteristics, with the structures in the vorticity and in the current behaving in the same way. We also study the effect of Reynolds number on cluster statistics, and we finally analyze the properties of these clusters in terms of their velocity-magnetic field correlation. Self-organized criticality features have been identified in the dissipative range of scales. A different scaling arises in the inertial range, which cannot be identified for the moment with a known self-organized criticality class consistent with MHD. We suggest that this range can be governed by turbulence dynamics as opposed to criticality, and propose an interpretation of intermittency in terms of propagation of local instabilities.
\end{abstract}
\maketitle

\section{Introduction}
As the resolving power of experimental instrumentation increases, turbulent flows as they occur in geophysics and astrophysics are being examined with more accuracy, and the multiplicity of scales in interactions becomes more apparent. To take an example, the modal distribution of energy in the Solar Wind has been known for a long time to follow a power law close to the Kolmogorov prediction for incompressible fluid turbulence \cite{voyager} (see \cite{SW} for reviews), although its physical environment is infinitely more complex than what was first envisaged by Kolmogorov, involving magnetic fields and coupling to acoustic and whistler modes, to name but a few phenomena at play. There are also numerous observations of spatially correlated turbulent structures and flows in Earth's magnetosphere. Recent observations from Cluster and THEMIS multi-spacecraft missions \cite{Cluster,weygand} provide a sophisticated physical picture of a variety of significant effects, for example, intermittent (spatially sparse) structures and transient plasma transport associated with reconnection in the tail plasma sheet and at the dayside magnetopause, formation of shocks and small-scale magnetic filaments, Kelvin-Helmoltz vortices and coherent structures viewed as Alfv\'enic turbulence \cite{recent_nature}, as well as other effects. Signatures of MHD turbulence are also found, e.g., in the magnetosphere of Jupiter \cite{saur}, and in the interstellar medium \cite{falga}. 

Of the many features now being resolved in the observations, intermittency in turbulence is of critical importance as it is linked with magnetic energy conversion and dissipation in solar-terrestrial plasmas. Two well-known examples of this link are flaring activity in the solar corona, and magnetospheric substorms in the tail plasma sheet of Earth's magnetosphere. In both cases, free magnetic energy is released through spatially localized reconfiguration of the plasma geometry which is significantly affected by MHD turbulence. Intermittent magnetic structures in the solar corona can generate multiple tangential discontinuities leading to energy avalanches and strongly inhomogeneous dissipation \cite{char01,ur06,dmit97}. An enhanced intermittency often reflects the formation of an unstable magnetic topology. The latter has been explored in detail when examining data from solar active regions which reveal precursory dynamics of intermittent measures prior to large solar flares \cite{abramenko07}. MHD intermittency is also likely to be a major factor defining initial locations of magnetic reconnection events in the nearly collisionless plasma of the Earth's magnetotail \cite{ang99,klimas,vor06}. It can be a triggering mechanism for a variety of instabilities of plasma behavior at both kinetic and MHD scales \cite{lui01, klimas} responsible for multiscale particle precipitation in the night-time auroral region \cite{vadim1, stepanova}. The timing, position, and energy output of such events -- as well as the structure of the Solar Wind mediating their interaction -- are largely unpredictable, reflecting the stochastic nature of the underlying fluid dynamics.

The intermittent structures associated with dissipation in turbulent flows are difficult to detect because they reside mostly at small scale, in thin current and vorticity sheets, and are entangled with ambient plasma flows in a topologically complex way. These structures are dynamically important as they participate in the local heating of the medium, for example through localized magnetic reconnection events, and are important not only in astrophysics but also in laboratory plasmas (see, e.g., \cite{gekelman}). Intense and localized dissipative structures in MHD flows have also been obtained numerically, in both two (2D) \cite{intermi} and three space dimensions (3D) \cite{third}. Intermittent dissipation in MHD simulations is found to be typically stronger than that in neutral fluids (i.e., with a stronger departure from self-similarity), and it can vary in time, as observed in solar active regions \cite{sorriso04,abramenko07}. A striking feature of these structures, as found in many simulations, is a high degree of correlation between the velocity and the magnetic field both globally \cite{vb} and locally \cite{servidio2}. However, the precise relationship between these structures and the global statistical properties of the flow is not well understood, although it is known that in 2D and in 3D the structures can interact with the underlying turbulent flow to affect properties such as the global dissipation through local processes like reconnection. Furthermore, at high Reynolds numbers these structures can have complex geometries, e.g., roll-up and fold as observed in the Solar Wind \cite{hasegawa} and in direct numerical simulations (DNS) \cite{1536b}, complicating significantly attempts to make a connection between structures and statistical properties.

Ensemble-based description of the geometry of intermittent dissipation is an important issue as a turbulent flow is characterized not only by the structures that develop within, but also by the statistical properties of the flow as a whole. For fluids in 2D, the statistics of vortices have been studied in detail \cite{mcwilliams} and a relationship between vorticity and stream function has been found which can be ascribed to a distribution of signed vortices \cite{matthaeus_2d} using a maximum entropy principle \cite{robert} (see also \cite{chavanis} for a recent analysis). In three dimensions, the situation is much more complex but in one specific case, a Kolmogorov spectrum for the energy has been obtained analytically from the dynamics of the stretching of a spiral vortex  \cite{lundgren}. However, while high-intensity dissipative structures in 3D MHD have been successfully studied for a number of years  \cite{vapor_id}, mostly through thresholding and visualization of current and vorticity (see, e.g., \cite{PPS95}--\cite{undul} in 3D, and \cite{servidio} for a thorough study of reconnection events in 2D), we are not aware of any ensemble-based studies of turbulent structures observed at intermediate to small intensity levels in high Reynolds number 3D MHD. A quantitative analysis, which requires the development of new software tools, is particularly important in the wake of two overarching developments: the emergence of petascale computers that will produce vast amount of data and detailed point-wise information about the relevant dynamical variables and their derivatives, as well as planned {\it in situ} observations, in particular those in association with the upcoming NASA's Magnetospheric Multiscale (MMS) mission, which will investigate the role of turbulence and other cross-scale phenomena in fast magnetic reconnection.
 
Therefore, in this paper we propose a new methodology for analyzing cross-scale behavior of three--dimensional MHD turbulence enabling the detection of multiple dissipative structures at arbitrary intensity levels. We use our tools to extract current and vorticity structures in numerical simulation outputs with two distinct types of initial conditions. The results obtained show the existence of robust scaling behavior in both the inertial and dissipative regimes of scales in the turbulent fluids we study. The reported scaling exponents shed new light on the role of the initial conditions and of the Reynolds number in the formation of intermittent dissipative structures in current and vorticity fields. Finally, our analysis supports the possibility of self-organized critical behavior for some of the small-scale structures we detect with our algorithm.

\section{Methodology}
\subsection{Equations and flows}

The incompressible MHD equations in dimensionless Alfv\'enic units and in the absence of forcing read:
\begin{eqnarray}
&& \partial_t {\bf v} + {\bf v} \cdot \nabla {\bf v} = 
    -\rho_0^{-1} \, \nabla {\cal P} + {\bf j} \times {\bf b} + 
    \nu \nabla^2 {\bf v} , 
\label{eq:MHDv} \\
&& \partial_t {\bf b} = \nabla \times ( {\bf v} \times
    {\bf b}) +\eta \nabla^2 {\bf b} \ ,
\label{eq:MHDb}
\end{eqnarray}
with ${\bf v}, \ {\bf b}$ being the velocity and the magnetic fields, ${\bf j}=\nabla \times {\bf b}$ the current density, ${\cal P}$ the pressure, $\rho_0$=1 the constant density, and ${\bf \nabla} \cdot {\bf v} = \nabla \cdot {\bf b} = 0$. When the viscosity $\nu$ and the magnetic resistivity $\eta$ are both equal to zero, the energy $E_T=\left<v^2+b^2\right>/2$, cross helicity $H_C=\left<{\bf v} \cdot {\bf b}\right>/2$, and magnetic helicity $H_M=\left<{\bf A} \cdot {\bf b}\right>$ (where ${\bf A}$ is the vector potential, ${\bf b} = \nabla \times {\bf A}$) are conserved. Equations (\ref{eq:MHDv})-(\ref{eq:MHDb}) have been solved in a three-dimensional box with periodic boundary conditions and a pseudospectral code dealiased by the 2/3 rule; $k_\textrm{min}=1$ for a box of length $L_0=2\pi$, and with $N$ regularly spaced grid points, this leads to a maximum wavenumber $k_\textrm{max}=N/3$. At all times, $k_D/k_\textrm{max}<1$, with $k_D$ denoting the dissipation wavenumber, in order to ensure an accurate numerical computation down to the smallest resolved scale. 

Two types of flows are studied in this paper, which were computed on regular grids ranging from $512^3$ to $1536^3$ points (see Table \ref{table:runs} for a brief presentation of the runs; see also \cite{Mininni06,second_fluid} where these runs are described in the context of a study of the general properties of MHD turbulence). In the first flow (runs I and II), the fields are constructed from a superposition of Arn'old-Beltrami-Childress (ABC) flows (see, e.g., \cite{ABC}), at wavenumbers $k=1$ to $k=3$, to which smaller-scale random fluctuations are added with a spectrum $k^{-3}\exp [-2(k/k_0)]^2$ for $k>3$ (see \cite{Mininni06}). The phases of the modes with $k>3$ are chosen from a Gaussian random number generator in such a way that the initial cross-correlation of the two fields is negligible: initially, $E_V = E_M = 0.5$, $H_C\approx 10^{-4}$, and $H_M =0.45$.

Another flow we compute (run III) is that of the so-called Orszag-Tang vortex (OT hereafter) generalized to MHD \cite{PPS95} (see also \cite{Mininni06} and references therein). This flow has been studied at length in two space dimensions for its reconnection properties (it has a magnetic X-point centered at a stagnation point of the velocity); its generalization to 3D is straightforward, with a simple sinusoidal variation in the  $z$ direction. Initially, $E_V=E_M=2$, 
$H_C= 0.41$ and $H_M=0$. Note that the two types of initial conditions differ in their invariants: the OT flow has zero magnetic helicity and a sizable cross correlation, whereas for the ABC flow, it is the opposite.

\subsection{Cluster detection}

\begin{table} \caption{
Nomenclature of runs with $N$ the linear grid resolution, and $\nu$ and $\eta$ the viscosity and magnetic diffusivity, respectively. The Reynolds number $R_V=U_0L_0/\nu$ and Taylor Reynolds number $R_\lambda= U_{0}\lambda /\nu$ (with $U_{0}$ the {\it r.m.s.} velocity, $L_0$ a characteristic large-scale and $\lambda=2\pi [E_T / \int k^2 E_T(k)dk ]^{1/2} $ the Taylor scale based on the total energy spectrum) are both evaluated at peak of dissipation.}
\begin{ruledtabular} \begin{tabular}{cccccc}
Run &Type  & N   & $\nu=\eta$                    & $R_V$  & $R_\lambda$   \\
\hline
I        & ABC & 512 & $6.    \times 10^{-4}$    & 3100   & \ 630    \\
II      & ABC    & 1536 & $2. \times 10^{-4}$   & 9200   & 1100    \\
III      & OT    & 512 & $7.5 \times 10^{-4}$   & 3300   & \ 880    
\label{table:runs} \end{tabular} \end{ruledtabular} \end{table}

Turbulent flows exhibit small-scale structures with strong gradients in the vicinity of which dissipation takes place. In principle, detection of structures can be done on any physical variables but more essential to a turbulent flow with its complex small-scale behavior are vorticity and current. Indeed, the primary channels of spatially localized energy dissipation in a resistive MHD fluid are the Joule heating, proportional to the squared current density $j^2 = |\nabla \times {\bf b}|^2$, and the kinetic dissipation that can be characterized indirectly by the squared vorticity $\omega^2= |\nabla \times {\bf v}|^2$ (note that the local dissipation of kinetic energy is proportional to the symmetric part of the velocity gradient matrix, the difference stemming from the fact that ${\bf v}$ is a vector whereas ${\bf b}$ is an axial vector).

Our analysis is thus focused on 3D arrays containing the values of $j^2$ and $\omega^2$ for each of the $N^3$ grid nodes of the simulations listed in Table \ref{table:runs}. A grid node is treated as belonging to a dissipative structure if the dissipation in this node, expressed in terms of either field, exceeds the level of $a_{th}$ standard deviations above the mean value, with  $a_{th} \in [1,3]$: 
\begin{eqnarray}
j^2_{th} &=& \left\langle j^2 \right\rangle + a_{th} \sqrt{\left\langle j^4 \right\rangle  - \left(\left\langle j^2 \right\rangle\right)^2 },
\label{eq:th1}\\
\omega^2_{th} &=& \left\langle \omega^2 \right\rangle + a_{th} \sqrt{\left\langle \omega^4 \right\rangle  - \left(\left\langle \omega^2 \right\rangle\right)^2, }
\label{eq:th2}
\end{eqnarray}
and where $\left\langle\cdot \right\rangle$ denotes averaging over the entire simulation volume as before. Intermittent dissipation structures in the $j^2$ field (current sheets) are defined as spatially connected sets of grid nodes satisfying the condition $j^2 > j^2_{th}$; the structures in the $\omega^2$ field (vorticity sheets) are defined similarly, based on $\omega^2 > \omega^2_{th}$. Our goal is to separate these structures from the background and to study their individual properties such as linear size, volume, or internal dissipation rate, for subsequent ensemble-based surveys. 

In order to overcome inherent memory limitations of standard cluster analysis algorithms such as, e.g., the {\it Label Region} function of the Interactive Data Language (IDL), we have developed a new technique enabling fast decomposition of multidimensional data arrays into sets of dissipative structures while dramatically reducing the amount of stored information. 

The first step of our technique consists of building a table of contiguous intervals (activation sites) along a chosen direction (``scanning direction'') where the studied data field exceeds the detection threshold. We tabulate boundaries of such intervals for all relevant positions in the $d-1$ coordinate space orthogonal to the scanning direction, $d$ being the dimensionality of the original data set. By design, the choice of the scanning direction is arbitrary and does not affect the detection results. The second step is to find and label spatially connected clusters of activations using the ``breadth-first search'' principle to avoid backtracking of search trees representing individual clusters. We find that it is important to consider all of the $3^d-1$ nearest neighbors of each grid node, including the diagonal neighbors, when identifying connected activations. Finally, the activation table is sorted a second time according to the cluster labels, in order to provide faster access to the detected structures. 

The output data array preserves complete information on the location and shape of all the contiguous regions in the simulation volume where the threshold condition is fulfilled. The size of this array is typically  smaller by two order of magnitudes (depending on the threshold), compared to the original data. For the purpose of this paper, the described technique is used to identify structures in static snapshots of a turbulent fluid. However, it can be easily extended to higher-dimensional data sets representing spatiotemporal dynamics of turbulent structures, with the time axis being a natural scanning direction.

\subsection{Data analysis tools}

\begin{figure}
\includegraphics[width=8.5cm]{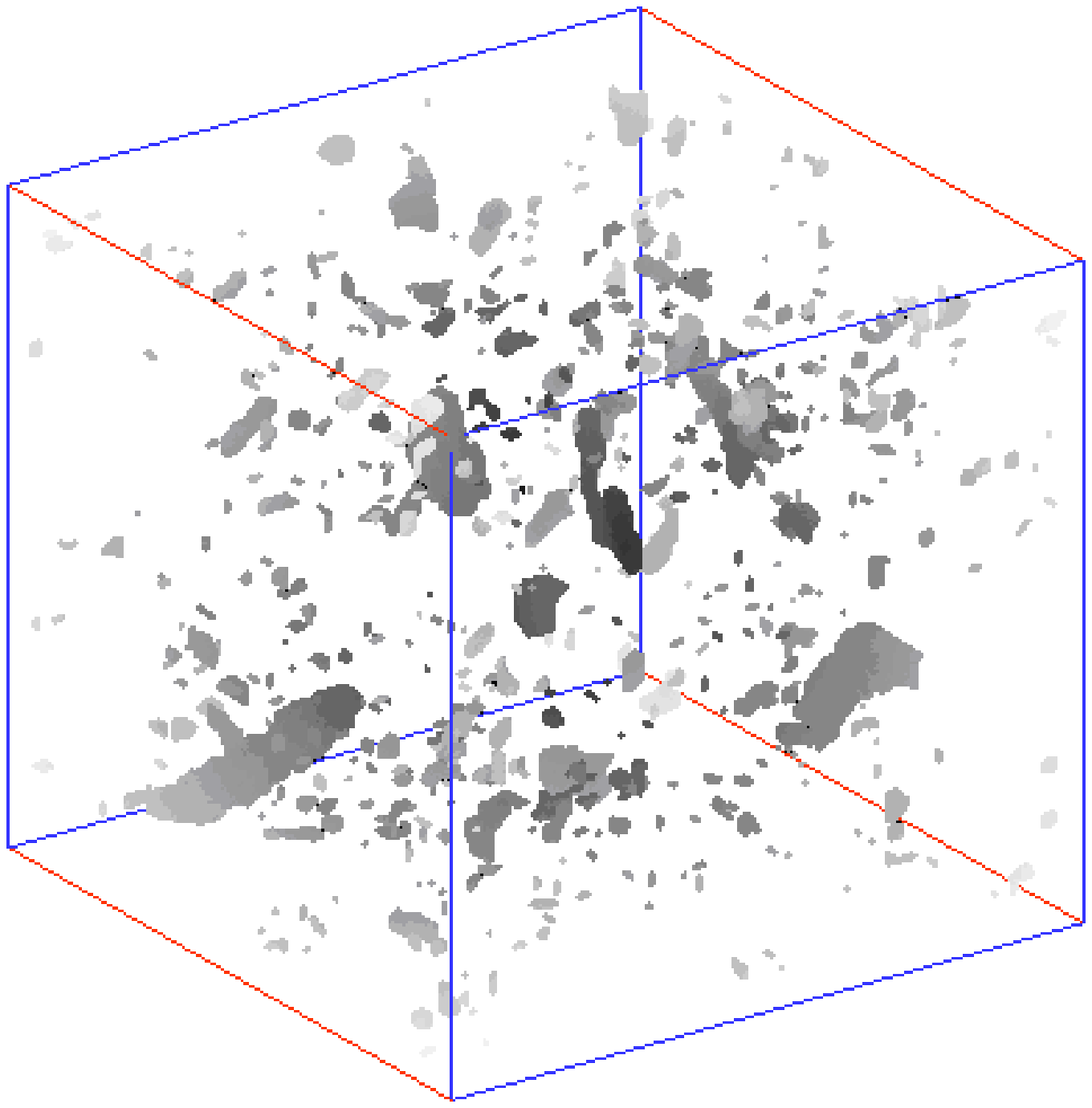}
\includegraphics[width=4.25cm]{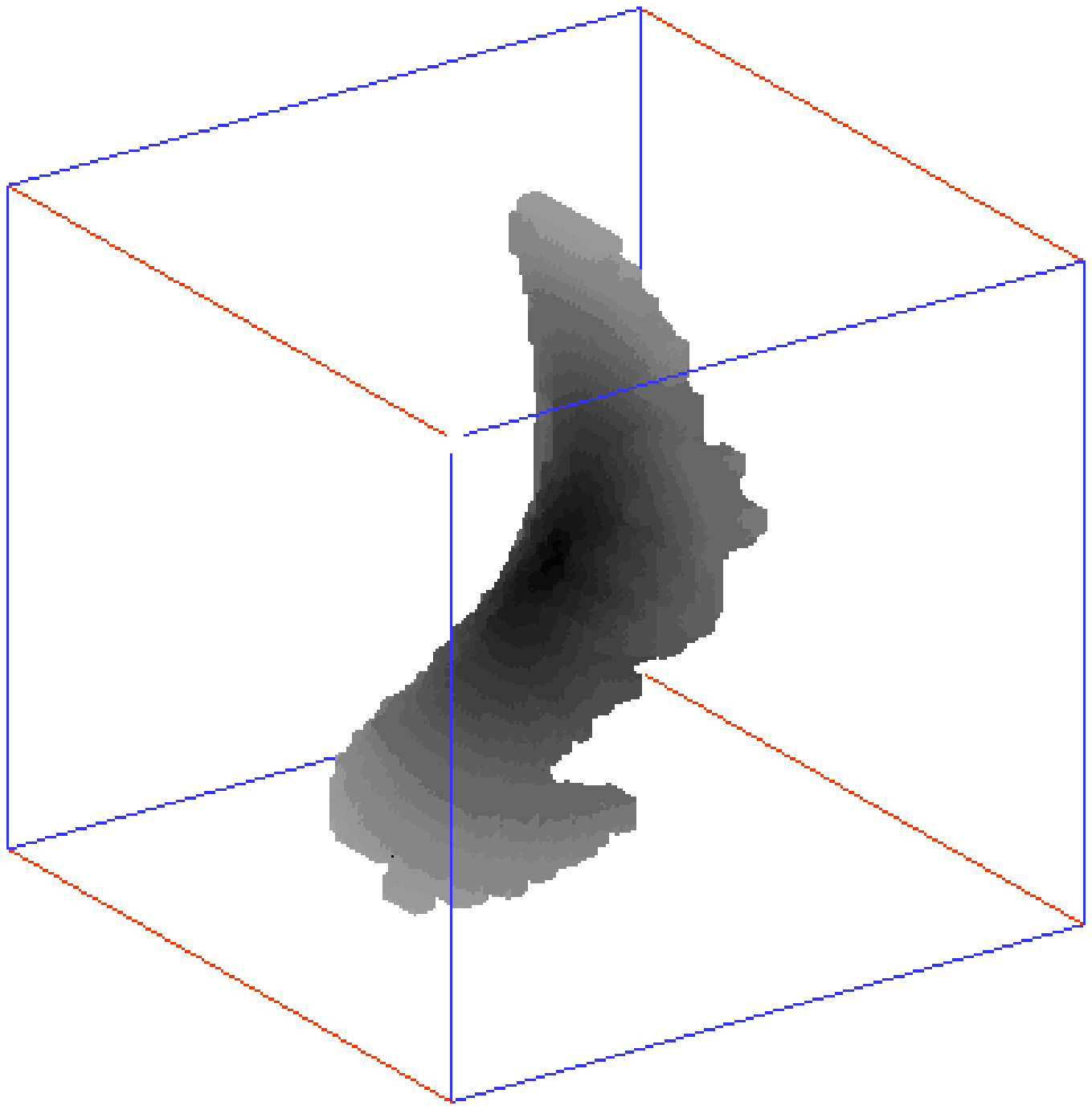}
\includegraphics[width=4.25cm]{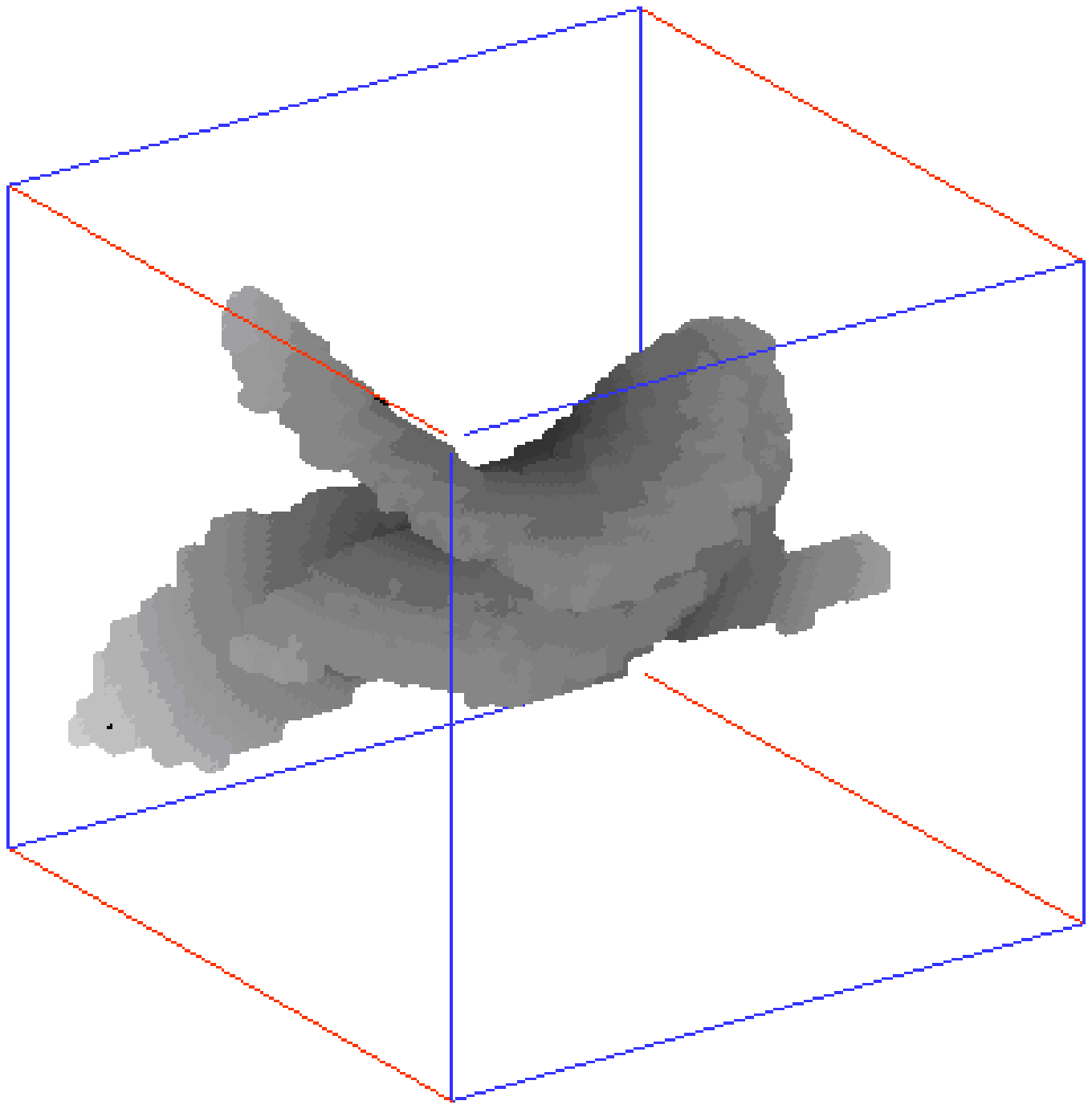}
\caption{({\it Color online})
{\it Top:} Global view of dissipative clusters in the $j^2$ field selected by our algorithm in the the ABC run I. The largest cluster has been removed to let see the intermediate size clusters; only one tenth of the remaining clusters are shown.
{\it Bottom:} Zoomed-in view of two selected current clusters, showing strong curvature of the sheets; the vorticity behaves similarly. The lower right edge (direction marked with red in the color version) is parallel to the $z$ axis chosen as the scanning direction.
} \label{f_curved} \end{figure}

\begin{figure}
\includegraphics[width=8.5cm]{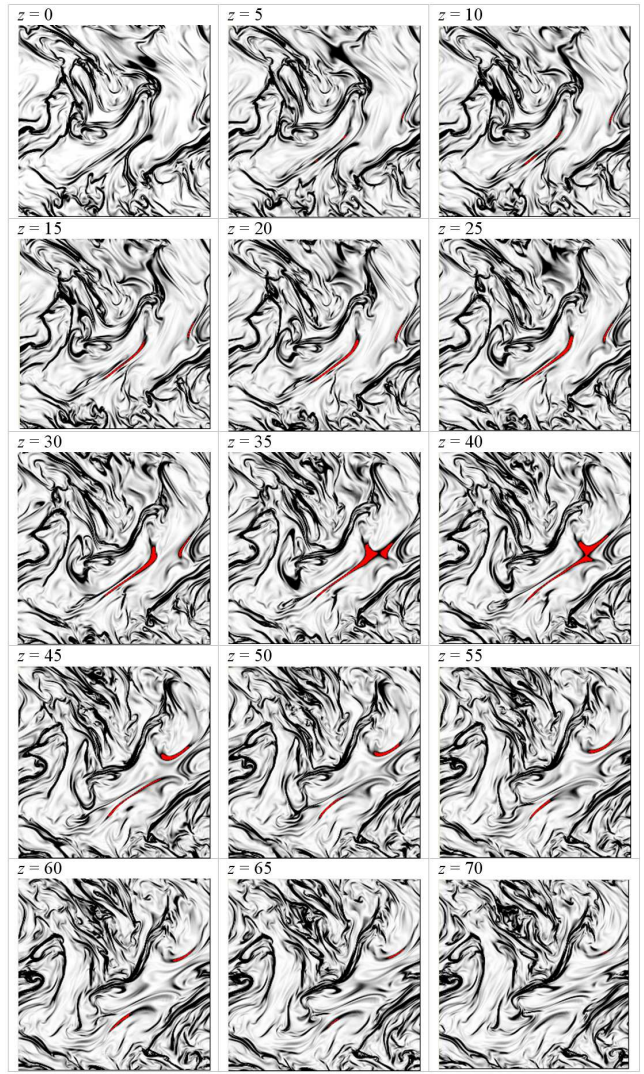}
\caption{({\it Color online})
Two-dimensional visualization of the current at fifteen regularly spaced slices in the $z$ direction with $z\in[0,70]$ (in pixel units). A dissipative cluster made up of two twisted current sheets merging is shown in the middle of the box (highlighted in red in the color version).
} \label{f_slice} \end{figure}

The data sets analyzed in this paper are three--dimensional cubes ($d=3$) described by $x$, $y$ and $z$ coordinates. The scanning direction is chosen parallel to the $z$ axis. The activation tables represent lists of $z$ intervals with $[j(x,y,z)]^2 > j^2_{th}$ or $[\omega(x,y,z)]^2 > \omega^2_{th}$, ordered according to their projected position onto the $x - y$ plane. 

The following primary parameters are computed and stored for each of the detected current and vorticity structures:
\begin{eqnarray}
L_i  &=& \delta \, \underset{k, l \in \Lambda_i}{\text{max}} |\, {\bf r}_k - {\bf r}_l \,|
\label{eq:par_first}\\
L_{\{x,y,z\}i} &=& \delta \, \underset{k, l \in \Lambda_i}{\text{max}} |\, \{x,y,z\}_k - \{x,y,z\}_l \, | \\
R_i  &=& \sqrt{L_{xi}^2+L_{yi}^2+L_{zi}^2} \\
V_i  &=&  \delta^3 \sum_{k\in \Lambda_i}{k} \\
A_i  &=&  \delta^2 \!\!\!\!\!\!\sum_{\substack{k\in \Lambda_i \\ M(k) \not\subset \Lambda_i}}\!\!\!\!\!\! k \\
P_{\{j,\omega\}i}  &=&  \delta^3 \left\{
\begin{array}{rl}
\displaystyle{
\sum_{k\in\Lambda_i}{j^2({\bf r}_k)},} \\
\displaystyle{
\sum_{k\in\Lambda_i}{\omega^2({\bf r}_k)} \ .}
\end{array} \right.
\label{eq:P} 
\end{eqnarray}
Here, $\Lambda_i$ is the set of all the grid node indices belonging to the $i$th dissipative structure, the subindices $k$ and $l$ label individual grid nodes, $M(k)$ is the full set of the 26 nearest neighbors of the $k$th node, ${\bf r}_k$ is the dimensionless position vector of the $k$th node, and finally $\delta=2\pi/N$ is the grid spacing, uniform and isotropic. The primary parameters of the clusters are then as follows. $L_i$ is the linear size of the structure defined as the largest pairwise distance (diameter) between the points tabulated in $\Lambda_i$, $L_{\{x,y,z\}i}$ are the maximum dimensions of the structure projected onto each spatial direction, $R_i$ is the characteristic linear scale of the smallest Euclidean volume embedding the whole structure, $V_i$ is the physical volume occupied by the structure, and $A_i$ is the area of its outer surface. $P_{\{j,\omega\}i}$ is the volume-integrated contribution of the structure to the kinetic and magnetic enstrophies, with the global corresponding quantities defined as $\Omega_j = \int j^2 dV$ and $\Omega_\omega = \int \omega^2 dV$ respectively, and where the integral is over the entire box. As mentioned before, except for a constant of proportionality ($\nu=\eta$) these quantities are also the dissipation rates, and in the following we refer to $P_{\{j,\omega\}i}$ as the ``kinetic and magnetic dissipation'' of the structure.

In addition to these primary parameters, the following measures characterizing the geometry of the structures were used:
\begin{eqnarray}
\chi_i &=&  \left\langle 
\frac{ {\bf v}( {\bf r}_k) \cdot {\bf b}({\bf r}_k) }{ |{\bf v}( {\bf r}_k)| \, |{\bf b}({\bf r}_k)| }
\right\rangle_{k\in \Lambda_i}
\\
H_i  &=&  V_i/(A_i/2)  \\
C_i &=& \pi (L_i/2)^2/(A_i/2) \ ,
\label{eq:par_last}
\end{eqnarray}
where $\chi_i$ is the cosine of the local alignment angle between the velocity and the magnetic field vectors averaged over all the positions within the $i$th structure, $H_i$ and $C_i$ are respectively the characteristic thickness of the structure and its topological complexity, both computed under the assumption of a sheet-like geometry which we validate later in the text. $C$ is defined as the ratio between the area of a circle with the diameter equal to the linear size $L_i$ of the structure, and the actual area of one of the sides of the structure. As follows from the definition, $C$ increases if the structure has holes or other irregularities reducing $A$ for a given $L$, and decreases if the structure has a curved shape ensuring a more efficient spatial filling for a fixed linear scale. The relative contribution of the second effect is expected to grow with $L$ (larger structures tend to roll-up and fold more frequently than the smaller ones), while the first effect is nearly scale-invariant as we show below.

Two major groups of statistics are invoked to quantify the scaling behavior of the detected structures. The first group includes a set of regression plots characterizing the geometric scaling of the structure parameters with respect to the linear size $L$; the second group is represented by a set of probability distribution functions of structure parameters. Both types of statistics are approximated by power-law dependencies:
\begin{eqnarray}
X(L) &\propto& L^{D_X}, \\
p(X) &\propto& X^{-\tau_X},
\label{eq:exp} \end{eqnarray}
in which $X$ denotes any of the parameters defined in Eqs. (\ref{eq:par_first})-(\ref{eq:par_last}), $p(X)$ is an estimated probability density function of $X$, and $D_X$ and $\tau_X$ are the geometric and the distribution scaling exponents, respectively; the latter are evaluated separately for the inertial and dissipative subranges of the flows we study. The ranges of linear scales corresponding to these subranges are determined based on the behavior of the energy spectra. For runs I and III, the inertial range scaling is observed for wave numbers $k \in [5, 30]$, an interval which corresponds to $L \in [0.21, 1.3]$. For run II, the inertial behavior is realized within the interval $k \in [5, 50]$ yielding $L \in [0.13, 1.3]$. For the dissipative (sub-inertial) scaling regime, we choose $L \in [0.025, 0.18]$ in runs I and III and $L \in [0.012, 0.11]$ in run II. 

The inertial and dissipative scaling ranges as specified in terms of $L$ are first applied to compute power-law fits describing the $X(L)$ statistics. Next, the fits are used to evaluate the inertial and the dissipative scaling ranges of the remaining parameters, and to estimate the distribution exponents $\tau_X$ corresponding to these ranges.

\section{Clusters and their properties}

\begin{figure}
\includegraphics[width=8.5cm]{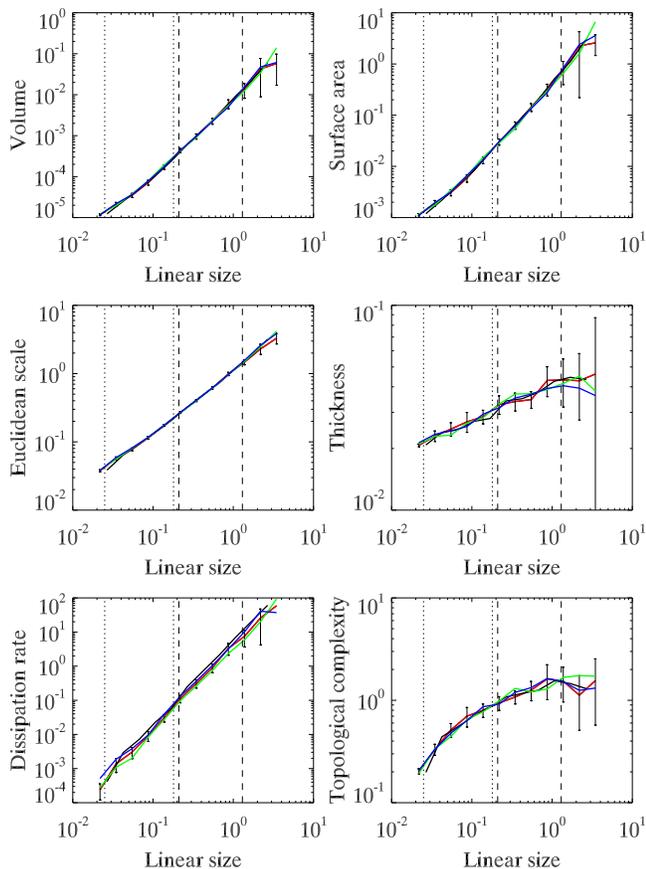}
\caption{({\it Color online})
Geometric scaling of the current sheet structures detected in run I based on the $j^2$ field, for several combinations of the snapshot time $t$ and of the threshold parameter $a_{th}$: $t=4$, $a_{th}=2$ (medium gray or red) and $a_{th}=3$ (black); $t=5$, $a_{th}=2$ (light gray or green) and $a_{th}=3$ (dark gray or blue). Dotted (dashed) vertical lines show the boundaries of the dissipative (inertial) scaling ranges used for computing the scaling exponents reported in Tables \ref{tab_inertial}--\ref{tab_comp}. The statistics are roughly insensitive to the detection threshold $a_{th}$ and are stable in time.}
\label{f_regr} \end{figure}

\begin{figure}
\includegraphics[width=8.5cm]{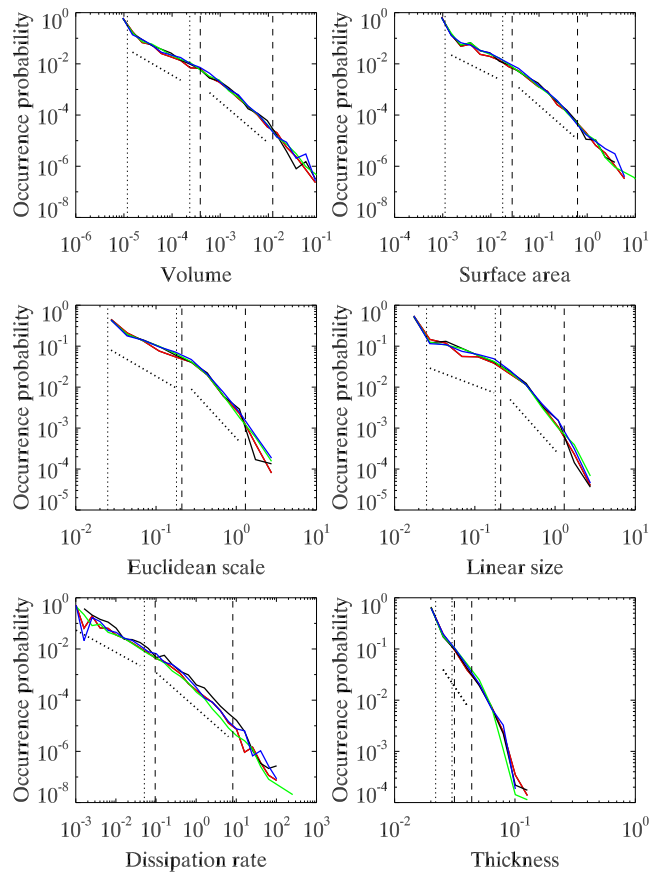}
\caption{({\it Color online})
Scaling of probability distributions of current sheet structures detected in run I. Color coding and notations are the same as in the previous figure. Tilted dotted lines show power laws in the inertial and dissipative ranges for $t=4$, $a_{th}=2$. As in the case of the geometric scaling, the shapes of the distributions are stable with respect to the detection threshold and time.}
\label{f_pdfs} \end{figure}

We now proceed to apply the algorithms described in the previous section to the data presented in Table \ref{table:runs}. We start by providing some qualitative examples illustrating the complexity of the intermittent turbulent structures under study, as well as  the performance of our cluster analysis code. Next, we present a detailed analysis of scaling behavior of these structures in the moderate-resolution runs I and III, followed by a comparative analysis of the same properties in the high-resolution run II. Our primary objectives will be to identify relevant parameters controlling the geometry of the observed structures, to clarify the role of the initial conditions, and to compare the geometry of the structures at inertial and dissipative scales. Finally, we discuss a possible link with self-organized criticality (SOC).

\subsection{The physical structures that emerge}

\begin{table} \caption{{\bf Inertial} range scaling exponents of current sheet structures in the runs with ABC and OT initial conditions (see Table \ref{table:runs}); length scales for analysis are $L \in [0.21, 1.30]$, $t=4$, $a_{th}=2$, $N=512$.}
\begin{ruledtabular} \begin{tabular}{lcccc}
 & Run I, $j^2$ & Run I, $\omega^2$ & Run III, $j^2$ & Run III, $\omega^2$ \\
\hline
\\
$D_L$ & 1.00 $\pm$ 0.00 & 1.00 $\pm$ 0.00 & 1.00 $\pm$ 0.00 & 1.00 $\pm$ 0.00  \\
$\tau_L$   & 2.20 $\pm$ 0.22 & 2.57 $\pm$ 0.22 & 1.86 $\pm$ 0.23 & 1.94 $\pm$ 0.23 \\
\\
$D_{R}$ & 0.94 $\pm$ 0.03 & 0.92 $\pm$ 0.01 & 0.93 $\pm$ 0.03 & 0.97 $\pm$ 0.02 \\  
$\tau_{R}$ & 2.14 $\pm$ 0.13 & 2.45 $\pm$ 0.24 & 1.90 $\pm$ 0.25 & 1.92 $\pm$ 0.21 \\
\\
$D_A$ & 1.71 $\pm$ 0.02 & 1.72 $\pm$ 0.07 & 1.72 $\pm$ 0.12 & 1.92 $\pm$ 0.05 \\
$\tau_A$ & 1.66 $\pm$ 0.08 & 1.81 $\pm$ 0.16 & 1.66 $\pm$ 0.11 & 1.39 $\pm$ 0.09 \\
\\
$D_V$  & 1.90 $\pm$ 0.06 & 1.90 $\pm$ 0.03 & 1.93 $\pm$ 0.12 & 2.15 $\pm$ 0.05 \\
$\tau_V$ & 1.61 $\pm$ 0.07 & 1.61 $\pm$ 0.06 & 1.49 $\pm$ 0.03 & 1.36 $\pm$ 0.10 \\
\\
$D_H$ & 0.18 $\pm$ 0.07 & 0.19 $\pm$ 0.04 & 0.21 $\pm$ 0.01 & 0.20 $\pm$ 0.05 \\
$\tau_H$ & 3.51 $\pm$ 0.13 & 3.78 $\pm$ 0.09 & 4.13 $\pm$ 2.0 & 7.02 $\pm$ 2.0 \\
\\
$D_P$ & 2.44 $\pm$ 0.10 & 2.32 $\pm$ 0.04 & 2.48 $\pm$ 0.17 & 2.96 $\pm$ 0.08 \\
$\tau_P$ & 1.44 $\pm$ 0.06 & 1.53 $\pm$ 0.05 & 1.30 $\pm$ 0.08 & 1.32 $\pm$ 0.06 \\
\\
\end{tabular} \end{ruledtabular}\label{tab_inertial} \end{table}

The top panel of Fig.~\ref{f_curved} gives a perspective view of specific examples among about 700 dissipative regions detected for the lower resolution ABC flow (run I) at $t=4$ and with $a_{th}=2$. The examples illustrate the complex multiscale nature of the $j^2$ dissipation field, a typical feature of turbulent fluids. The current structures can be as large as the whole grid (not included in the figure to make smaller ones visible), or as small as several grid spacings.

It should be emphasized that the upper panel in  Fig.~\ref{f_curved} is not produced by color-coding a continuous field as done often in turbulence visualizations. Each of the structures was first extracted by  the algorithm described in the preceding section. After that, roughly 1/10 of the structures were ``placed back'' in the domain according to their original positions and spatial orientations. We skipped the rest in order not to overcrowd the resulting picture.

The apparent two-dimensional geometry of the structures is typical for MHD turbulence, and it can be observed reliably over the entire inertial range of scales as we show in the next subsection. For smaller scales, the 2D geometry becomes questionable, partly because the current sheets tend to fold or roll to form tubes which can no longer be resolved.

The bottom panel of the same figure presents two typical examples of large-scale dissipative regions extracted by our code, each occupying about 2000 grid nodes. As one can see, the regions may have rather complicated overall shapes associated with twisting and splitting of current sheets. Since the code does not rely on any {\it a priory} for the cluster shape or size, it can efficiently identify both simple and complex (e.g., folded or rolled-up) current sheets across the entire range of relevant spatial scales. 

Figure \ref{f_slice} further illustrates the ability of the code to detect complex dissipative structures. Here, the gray background field represents the spatial distribution of $j^2$ in $x$-$y$ cross-sections of the data cube, with black (white) colors corresponding to the largest (smallest) current magnitude. The structures shown in the middle of the box (in red in the color version) correspond to a large-scale current sheet identified by our algorithm and embedded again into the turbulent flow to demonstrate its consistency with the surrounding MHD environment. In most of the slices, the highlighted current sheet consists of two disconnected pieces which only merge within a limited range of $z$ values. Despite this topological complexity, the structure has been correctly identified as a single set of contiguous grid nodes.

\subsection{Statistics of structures}

\begin{table} 
\caption{{\bf Dissipative} range scaling exponents of current sheet structures in the runs with ABC and OT initial conditions (length scales $L \in [0.025, 0.18]$, $t=4$, $a_{th}=2$,  $N=512$).}
\begin{ruledtabular}
\begin{tabular}{lcccc}
 & Run I, $j^2$ &  Run I, $\omega^2$ & Run III, $j^2$ & Run III, $\omega^2$ \\
\hline
\\
$D_L$ & 1.00 $\pm$ 0.00 & 1.00 $\pm$ 0.00 & 1.00 $\pm$ 0.00 & 1.00 $\pm$ 0.00  \\
$\tau_L$  & 0.75 $\pm$ 0.12  & 0.82 $\pm$ 0.04  & 1.31 $\pm$ 0.21  &  1.16 $\pm$ 0.08 \\
\\
$D_{R}$   &  0.81 $\pm$ 0.04 &  0.84 $\pm$ 0.05 & 0.89 $\pm$ 0.02  &  0.82 $\pm$ 0.03 \\
$\tau_{R}$   &  1.14 $\pm$ 0.09 &  1.22 $\pm$ 0.15 & 1.79 $\pm$ 0.09  &  1.65 $\pm$ 0.19 \\
\\
$D_A$   &  1.40 $\pm$ 0.11 & 1.44 $\pm$ 0.14  & 1.53 $\pm$ 0.08  & 1.29 $\pm$ 0.13  \\
$\tau_A$   & 0.95 $\pm$ 0.15  & 1.04 $\pm$ 0.10  & 1.38 $\pm$ 0.12  &  1.29 $\pm$ 0.14 \\
\\
$D_V$    & 1.51 $\pm$ 0.13  &  1.54 $\pm$ 0.15 & 1.60 $\pm$ 0.09  & 1.40 $\pm$ 0.15  \\
$\tau_V$   &  1.10 $\pm$ 0.12 & 1.16 $\pm$ 0.07  &  1.39 $\pm$ 0.10 & 1.40 $\pm$ 0.15  \\
\\
$D_H$   & 0.16 $\pm$ 0.01  & 0.15 $\pm$ 0.01  &  0.11 $\pm$ 0.02 &  0.11 $\pm$ 0.02 \\
$\tau_H$   &  3.51 $\pm$ 0.13 & 3.78 $\pm$ 0.09  & 4.13 $\pm$ 2.0  & 7.02 $\pm$ 2.0  \\
\\
$D_P$   & 2.26 $\pm$ 0.14  & 2.31 $\pm$ 0.13  & 2.32 $\pm$ 0.08  &  2.21 $\pm$ 0.24 \\
$\tau_P$   & 0.98 $\pm$ 0.08  & 1.13 $\pm$ 0.08  & 1.37 $\pm$ 0.08  & 1.33 $\pm$ 0.08  \\
\\
\end{tabular} \end{ruledtabular} \label{tab_dissip} \end{table}

Figure \ref{f_regr} shows the geometric scaling (dependence on length $L$, see Eq.~(\ref{eq:par_first})) of the volume $V$, area $A$, Euclidian scale $R$, thickness $H$, dissipation rate $P_j$, and the complexity $C$ on intermittent dissipative structures in the $j^2$ field of run I. Figure \ref{f_pdfs} shows the probability distributions of the same parameters, except for $C$; the shade of gray (colors) used in the figures represent four different combinations of the time of the snapshots $t$ (with $t=4$ or $t=5$, in units of the turn-over time of the problem, see \cite{Mininni06, third}), and of the threshold $a_{th}$ (with $a_{th}=2$ or $a_{th}=3$). The results obtained for these parameters as well as for $a_{th}=1$ (not shown) are indistinguishable within statistical uncertainty. Therefore, the scaling properties reported in this paper are not sensitive to the detection threshold, at least for $a_{th} \in [1, 3]$, and they do not vary on a time scale of the order of the turnover time. The dissipative and inertial ranges of scales are shown in both figures with vertical dotted and dashed lines respectively. 

The geometric scaling of the parameters $V$, $A$, $R$, and $P$ exhibits clear-cut power-law behavior, with the log-log slopes (the $D$ exponents of Eq.~(\ref{eq:exp})) undergoing slight changes at the transition between the two scaling regimes. The inertial values of $D_V$ and $D_A$ suggest that the studied structures have a nearly two dimensional geometry. This is an expected result since the co-dimension of MHD turbulence is equal to unity, suggesting sheet-like structures embedded in three-dimensional space \cite{PPS95, MB}. Similar $D_A$ estimates indicating sheet-like dissipative structures were obtained for the $\omega$ field in the same simulation run. A more careful inspection of the exponents obtained  herein shows that the inertial range values of both $D_A$ and $D_V$ are systematically below 2, suggesting that the structures have irregular edges. This interpretation is consistent with recently detected undulations of current sheet edges in the OT turbulence in 3D \cite{undul}.

Unlike the geometric scaling of other size measures, the thickness $H$ of the current sheets (Fig. \ref{f_regr}) does not vary significantly with $L$. It seems to saturate at the largest inertial scales revealing the existence of a characteristic thickness of the structures of $H =0.04 - 0.05$ (about 3-4 grid spacings). This thickness is likely representative of the turbulent dissipation length $\ell_{diss}$ whose estimated value is slightly larger than $3\times2 \pi /512 \approx 0.04$.

The geometric scaling of the topological complexity $C$ also demonstrates a saturation in the inertial regime, and is clearly non-scale free at smaller scales. As already mentioned in section II.C, the monotonic growth of $C$ across a range of scales can be attributed to an increasingly complex shape of the structures, and is also affected by their folding. 

The distribution functions (Fig. \ref{f_pdfs}) demonstrate  pronounced crossovers at the transition between the inertial and dissipative regimes. Overall, these crossovers are more evident than the crossovers in the $X(L)$ statistics shown in the previous figure. The thickness distribution is rather steep. It is likely to follow an exponential rather than a power-law decay, which is consistent with the existence of a characteristic thickness $H$ as discussed above. 

Tables \ref{tab_inertial} and \ref{tab_dissip} summarize inertial and dissipative range scaling exponents for runs I and III, using $a_{th}=2$, at $t=4$, corresponding to the maximum of dissipation. The first column in each table refers to the log-log slopes of the red curves in Fig. \ref{f_pdfs} and Fig. \ref{high_Re_regr}. The exponent values reported in Table \ref{tab_inertial} confirm that the geometry of both $j^2$ and $\omega^2$ structures observed in the inertial range is close to being two-dimensional. However, the dissipative range scaling (Table \ref{tab_dissip}) is significantly different, with the volume and area geometric exponents $D_V$ and $D_A$ being close to $1.5$, hinting at a fractal geometry of the structures with possible local anisotropy. 

The $\tau$ exponents characterizing $L$ (linear size) and $R$ (Euclidian scale) are almost identical in the inertial range (Table \ref{tab_inertial}) showing that either parameter can be invoked as a measure of linear scale. On average, the identity $L=R$ implies the absence of a preferred current or vorticity sheet orientation, indicating that at these scales, the MHD flows examined here are globally isotropic. The global isotropy obtains within the inertial range of scales and is not preserved at smaller scales, in accordance with previous results based on incompressible decaying MHD turbulence using ABC flows \cite{1536b}. Our analysis extends in an independent way these earlier findings demonstrating the inertial range global isotropy in both the OT and ABC runs.

Note the values of $\tau_H$ shown in Tables \ref{tab_inertial} and \ref{tab_dissip} are equal. This is a result of the steep non-power law decay of the thickness distribution, which prevented us from measuring this exponent separately in the dissipative and inertial ranges of scales.

When comparing the inertial distribution exponents characterizing the size of the structures, one can notice that these exponents are systematically lower in run III (OT initial conditions) than in run I (ABC initial conditions). This difference is not large but is statistically significant for several exponent pairs. Alternatively, the dissipative range scaling shows the opposite effect (OT distribution exponents are higher than the ones in the ABC run). Since the dissipation range exponents are smaller on average than the inertial exponents, the inertial/dissipative distribution crossover is more pronounced in run I as compared to run III; on the other hand, this makes the statistics of OT structures closer to scale-invariant across the entire range of the scales, as evidenced from the analysis of $D_P$, $D_V$, $\tau_P$, and $\tau_V$ estimates shown in Tables \ref{tab_inertial} and \ref{tab_dissip}. This may come from the fact that the OT flow has a well-defined structure with both partial zeros of the magnetic field (canceling of two components) and global zeroes (${\bf b}\equiv 0$), leading to more ordered reconnection events and turbulence developing at later times \cite{PPS95}, whereas the ABC runs have some random noise added at small scales which leads to more wrinkled structures of lesser extent.

To check the consistency of the pairs of $D$ and $\tau$ values we obtained, we also computed the exponents $\alpha_X=D_X(\tau_X-1)$ which should be equal for all scale-invariant measures $X$ of the examined sets of turbulent structures (due to the conservation of probability). We found the $\alpha_X$ values to be approximately constant for most of the structure parameters, which confirms the validity of our measurements. The only noticeable exception is $\alpha_H$ whose value is inconsistent with the other $\alpha$ exponents. This can be seen as additional evidence of a non-power law scaling for the thickness of the dissipative structures, likely with a well-defined characteristic scale. 

Table \ref{tab_comp} provides additional insight into the relationship between the various scaling exponents describing the parameters of $j^2$ and $\omega^2$ structures in the OT and ABC runs. The table shows estimated values of $\tau_{LX}$ for linear size distribution exponents (based on the probability conservation) compared to the exponents $\tau_L$ evaluated directly from $p(L)$ distributions. As one can notice, the values of $\tau_{LX}$ and $\tau_L$ are in a reasonable agreement for the inertial range, but different by a roughly constant factor for the dissipative range. Also, the inertial range estimates tend to be lower for run III relative to run I; for the dissipative range they are higher. The difference between the exponents corresponding to the two scaling ranges seems to be less pronounced for the OT run. As an example, compare $\tau_{LV} = 2.2$ ($1.2$) in the inertial (dissipative) regimes of the ABC flow with the values $\tau_{TV} =1.8$ ($1.6$) describing the same ranges in the OT flow. Therefore, the cross-over behavior in the linear size scaling is more evident for the ABC flow, in agreement with our conclusion based on the results in Tables \ref{tab_inertial} and \ref{tab_dissip}.

This lack of complete universality in scaling of MHD flows can be related to similar findings in different contexts. For example, it was shown in \cite{lee} that different power-law scaling for energy spectra can emerge with different initial conditions for MHD flows having the same invariants ($E_T$, $H_C$ and $H_M$ and $E_V=E_M$ at $t=0$). Different energy spectra have also been observed in the presence of an imposed uniform and strong magnetic field \cite{dmitruk,grappin,mason}.

\begin{table} 
\caption{Comparison of linear size distribution exponents evaluated using the relation $\tau_{LX}=D_X(\tau_X-1)+1$, $X \in \{R, A, V, P\}$
(see Eqs. (\ref{eq:par_first})-(\ref{eq:P})), with $\tau_L$ exponents obtained directly from $p(L)$ distributions ($t=4$, $a_{th}=2$, $N=512$) for ABC (run I) and OT (run III).}
\begin{ruledtabular}
\begin{tabular}{lcccc}
 & Run I, $j^2$ &  Run I, $\omega^2$ & Run III, $j^2$ & Run III, $\omega^2$ \\
\hline
\\
\multicolumn{5}{l}{\it Inertial range} \\
$\tau_{LR}$ & 2.07 & 2.34 & 1.84 & 1.89 \\
$\tau_{LA}$ & 2.13 & 2.40 & 2.14 & 1.75 \\
$\tau_{LV}$ & 2.16 & 2.15 & 1.94 & 1.78 \\
$\tau_{LP}$ & 2.08 & 2.24 & 1.74 & 1.95 \\
\\
$\left\langle \tau_{LX} \right\rangle$ & 2.11 $\pm$ 0.04 & 2.28 $\pm$ 0.11 & 1.92 $\pm$ 0.17 & 1.84 $\pm$ 0.09 \\
$\tau_L$  & 2.20 $\pm$ 0.22 & 2.57 $\pm$ 0.22 & 1.86 $\pm$ 0.23 & 1.94 $\pm$ 0.23 \\
\\
\multicolumn{5}{l}{\it Dissipative range} \\
$\tau_{LR}$   & 1.11 & 1.19  &  1.70 &  1.54 \\
$\tau_{LA}$   & 0.92 &  1.06 &  1.59 & 1.38  \\
$\tau_{LV}$   & 1.15  &  1.25 &  1.63 &  1.56 \\
$\tau_{LP}$   & 0.96  &  1.30 & 1.86  &  1.73 \\
\\
$\left\langle \tau_{LX} \right\rangle$  & 1.04 $\pm$ 0.11 & 1.20 $\pm$ 0.10 & 1.70 $\pm$ 0.12 & 1.55 $\pm$ 0.14 \\
$\tau_L$  & 0.75 $\pm$ 0.12  & 0.82 $\pm$ 0.04  & 1.31 $\pm$ 0.21  &  1.16 $\pm$ 0.08 \\
\end{tabular} \end{ruledtabular} \label{tab_comp}\end{table}

\subsection{Analysis at higher Reynolds number}

\begin{figure}
\includegraphics[width=8.5cm]{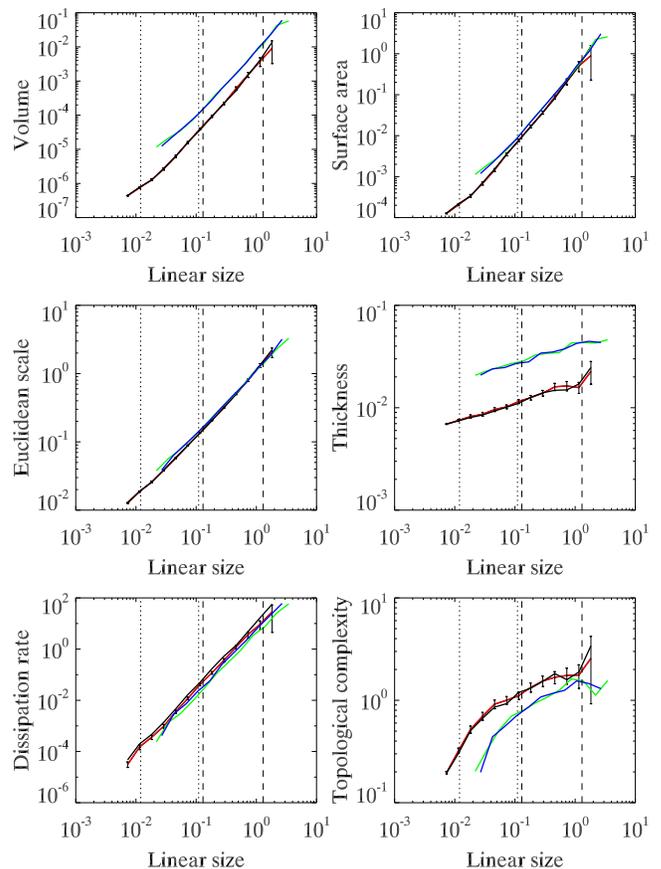}
\caption{({\it Color online})
Geometric scaling of current structures in runs I and II detected by thresholding the $j^2$ field at two different levels: $N=1536$, $a_{th}=2$ (medium gray or red lines) and $a_{th}=3$ (black); $N=512$, $a_{th}=2$ (light gray or green) and $a_{th}=3$ (dark gray or blue). Dotted (dashed) vertical lines show the boundaries of the dissipative (inertial) scaling ranges. The higher Reynolds number ABC flow (run II) develops considerably thinner current sheet structures described by smaller volumes, and roughly the same surface areas compared to the lower resolution run. The dissipation rate in run II is slightly higher, and the geometry of current sheets generated in this run is significantly more complex than the $j^2$ structures observed in run I.
}
\label{high_Re_regr} \end{figure}

\begin{figure}
\includegraphics[width=8.5cm]{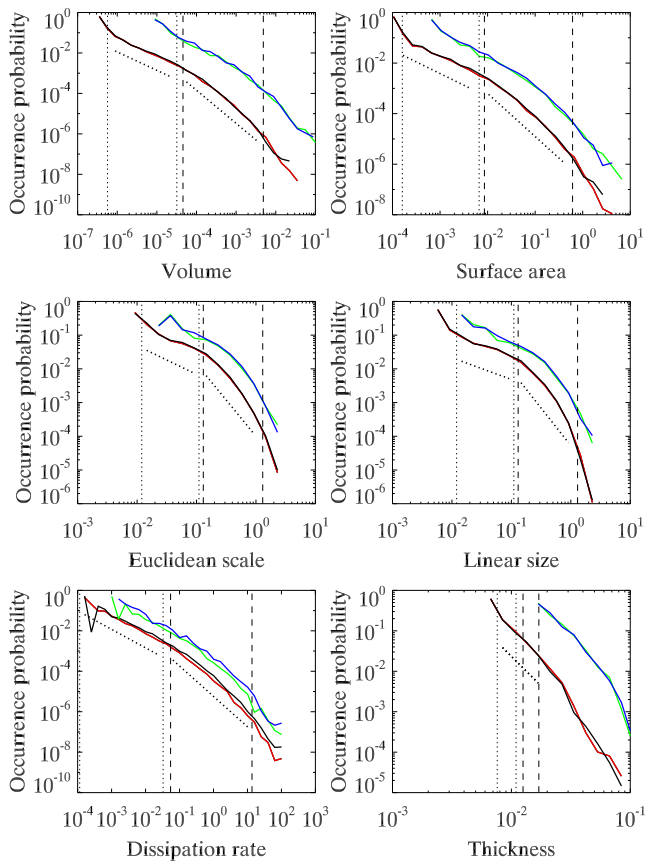}
\caption{({\it Color online})
Probability distributions of current structures in runs I and II. Tilted dotted lines show log-log regression slopes for run II, the remaining notations are the same as in the previous figure. With the exception of thickness distribution, all the statistics here exhibit power-law scaling with consistent sets of inertial range $\tau$ exponents.}
\label{high_Re_pdfs} \end{figure}

\begin{figure}
\includegraphics[width=8.5cm]{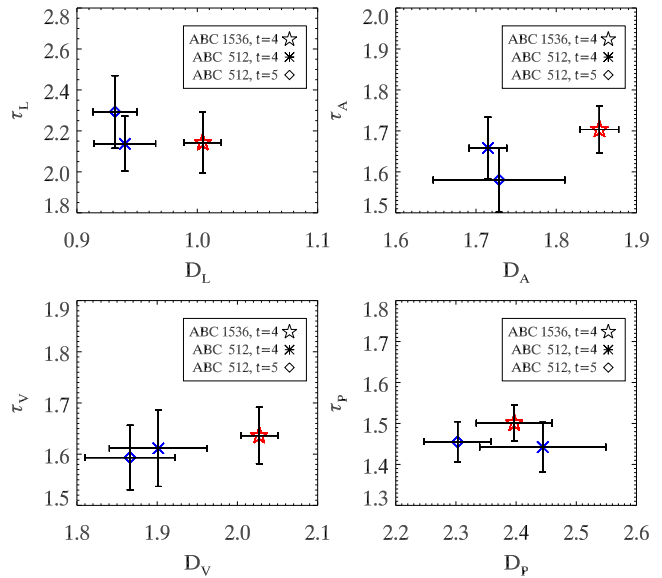}
\caption{({\it Color online})
Comparison of scaling exponents of current structures in runs I and II ($N=512$ and $N=1536$, respectively) detected by thresholding the $j^2$ field at the level of two standard deviations above the mean. The error bars are approximate confidence intervals ($\pm$ three standard errors) for each exponent. The plots show consistent distribution exponents but greater geometric exponents ($D_L$, $D_A$ and $D_V$) at higher Reynolds number. This difference becomes less prominent at higher detection thresholds (not shown). Similar tendencies are observed for the vorticity structures.}
\label{high_Re_scaling} \end{figure}

In order to determine what possible role the Reynolds number plays in the statistics of the structures studied in the preceding section, we now present an analysis of one snapshot for run II computed on a grid of $1536^3$ points and taken at peak of dissipation, and contrast it with run I (the runs have Taylor Reynolds number of $1100$ and $630$, respectively). The high resolution run ($N=1536$) is characterized by a larger Reynolds number and is therefore expected to generate more complex current and vorticity structures. A simple visual inspection of spatial patterns in $j^2$ and $\omega^2$ confirms this, and our quantitative analysis provides useful clues on the nature of the increased complexity in the high-resolution run.

Figures \ref{high_Re_regr} and \ref{high_Re_pdfs} show comparative statistics of runs I and II for the $j^2$ field. We expect, based on an approximate convergence of $j^2$ and $\omega^2$ scaling exponents in the lower resolution runs (see Tables \ref{tab_inertial}-\ref{tab_comp}), that the vorticity structures have a similar dependence on $N$. Note that the boundaries of inertial and dissipative ranges indicated by vertical lines in Figs. \ref{high_Re_regr} and \ref{high_Re_pdfs} are computed for $N=1536$ and are thus different from the boundaries in the previous figures at lower Reynolds number. 

The geometric scaling (Fig. \ref{high_Re_regr}) reveals a major distinctive feature of run II: it has measurably thinner dissipative structures, as expected if we associate $H$ with the dissipative scale $\ell_{diss}$ for this new run. Indeed, the volume of these structures is approximately half an order of magnitude smaller than the $V$ estimates in run I made at the same linear scale $L$. At the same time, the scaling of the area, likely dependent on the integral scale of the flow $L_{int}$, is remarkably similar. The discrepancy between runs I and II has a straightforward explanation, namely, significantly thinner structures in the high resolution ABC run. On average, the values of $H$ in this run are about three times smaller than the corresponding values in run I (for the same $L$). This difference is in agreement with the gain in the Reynolds number achieved due to the increase of the grid size from $N=512$ to $1536$ (or the decrease of the viscosity, see Table \ref{table:runs}). As in the case of run I, the dimensionless thickness of the smallest inertial structures in run II is about 3-5 grid nodes. It is interesting that the scaling of the dissipation rate corresponding to  structures for a higher Reynolds number flow is approximately the same as in run I, despite a significantly smaller volume and thickness of these structures. Therefore, the current sheets generated on the grid with $N=1536^3$ points are more intense (by a factor of $\approx 3$ in terms of the current, using the $H(L)$ scaling in the inertial regime). This is consistent with the finding that there is a finite dissipation rate in MHD in 3D \cite{third}, a fact well-known in the 2D case \cite{biskamp, politano89} and related to the possibility of fast reconnection in MHD even in the absence of a Hall current (see \cite{lapenta} and references therein for recent developments). Finally, the topological complexity of the structures in the high-resolution run is roughly twice the value characterizing the structures of run I, with $C=1$ matching the transition between the inertial and dissipative regimes. A similar match was found for $N=512$ (see Fig. \ref{f_regr}).

The probability distributions of $j^2$ structures in runs I and II (Fig. \ref{high_Re_pdfs}) show that in spite of the essential difference in the current sheet thickness, energy density and topological complexity evident from the geometric scaling, the probabilistic essence of the two runs is in fact quite similar. To a first approximation, the shape of the distributions is not influenced by Reynolds number, at least for the small range examined here. As in the case of the lower Reynolds number simulation at lower resolution, the distributions in run II exhibit well-defined "breaks" coinciding with the transition between the two scaling regimes (and again shifted toward smaller scales for the high $R_V$ run). It also shows a rapid decay of the probability distribution $p(H)$, consistent with the equivalent distribution in run I.

\begin{figure}
\includegraphics[width=8.cm, clip]{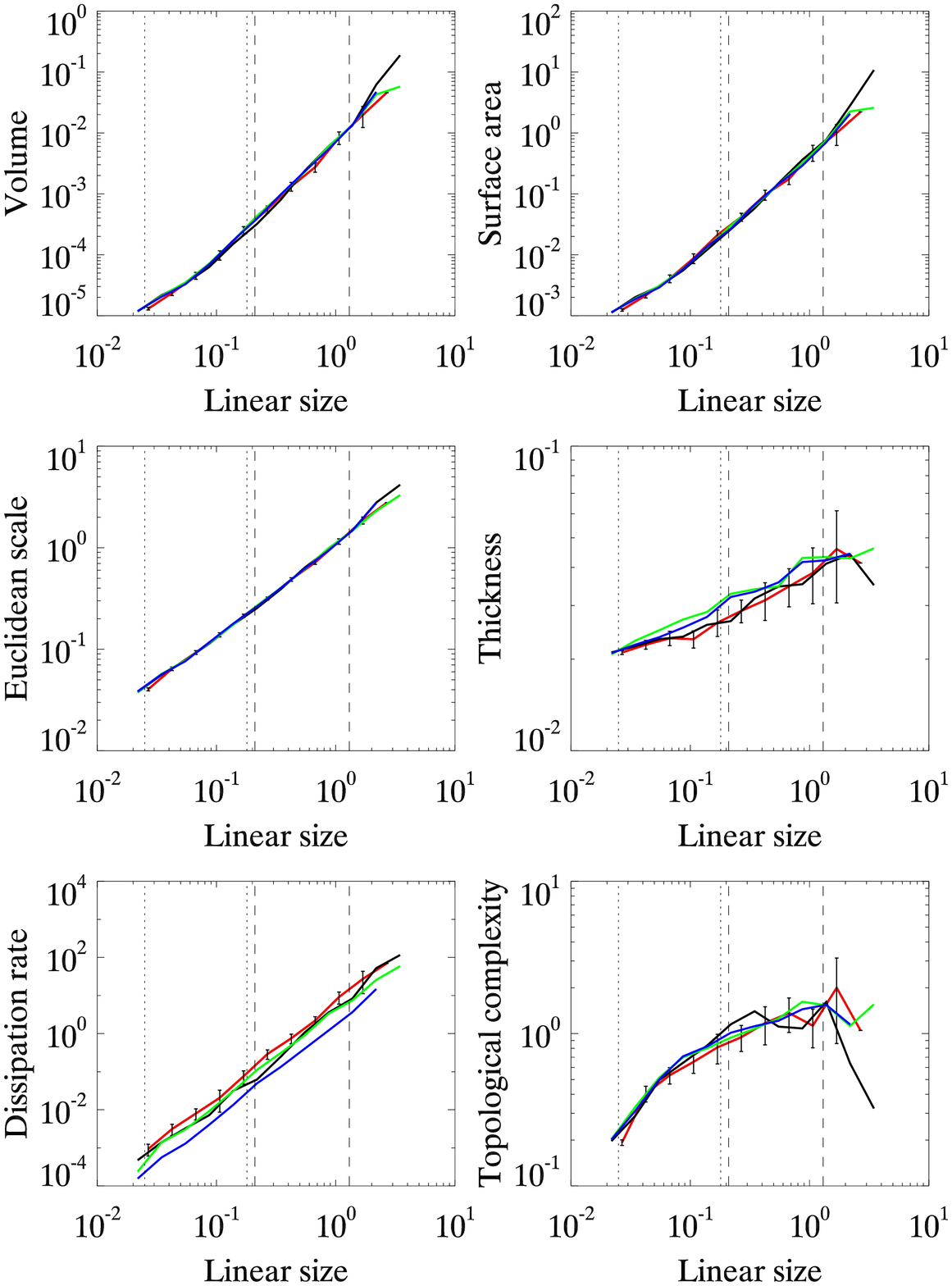}
\includegraphics[width=8.cm, clip]{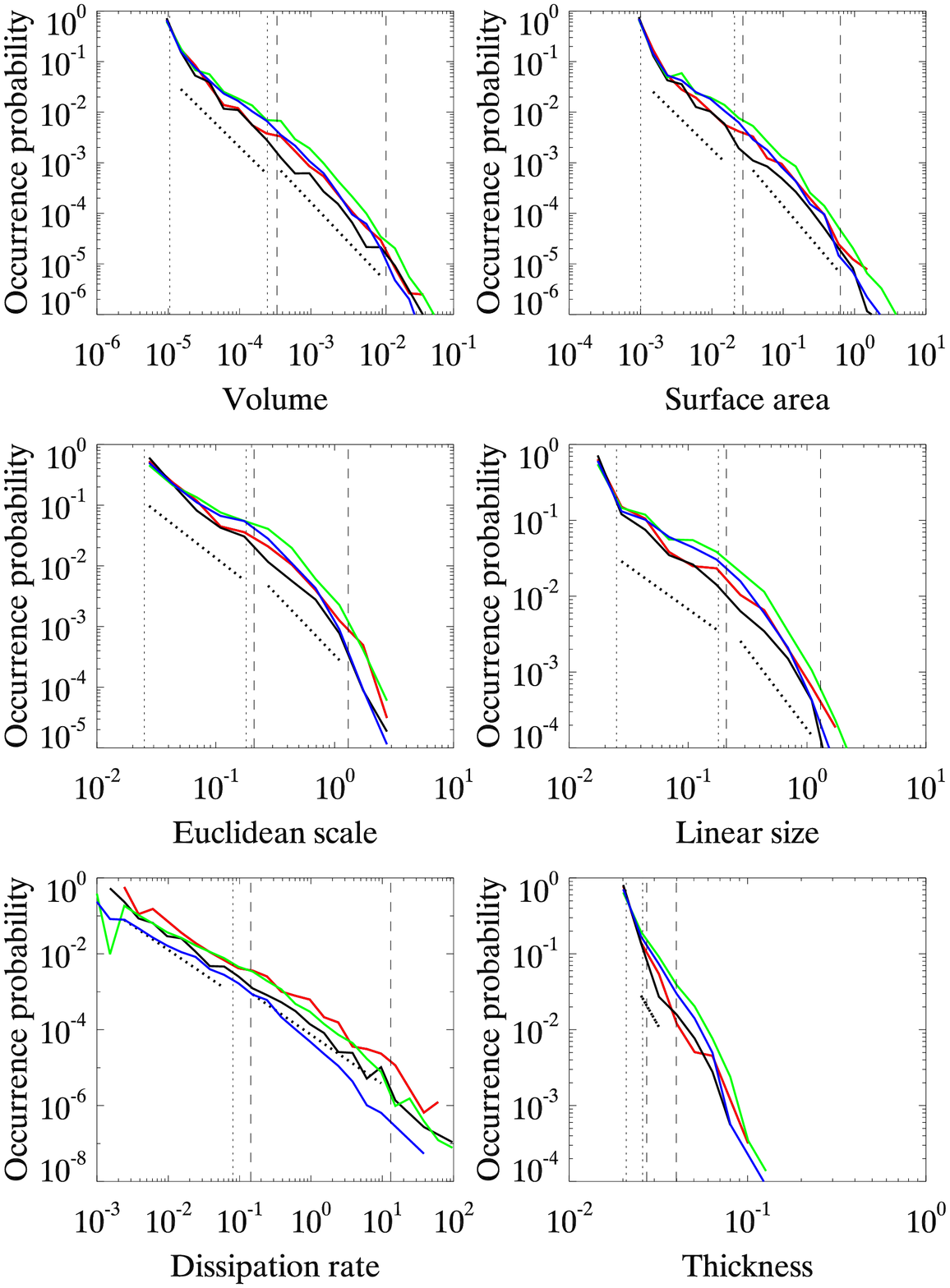}
\vskip 0.2 cm
\includegraphics[width=4.1cm, clip]{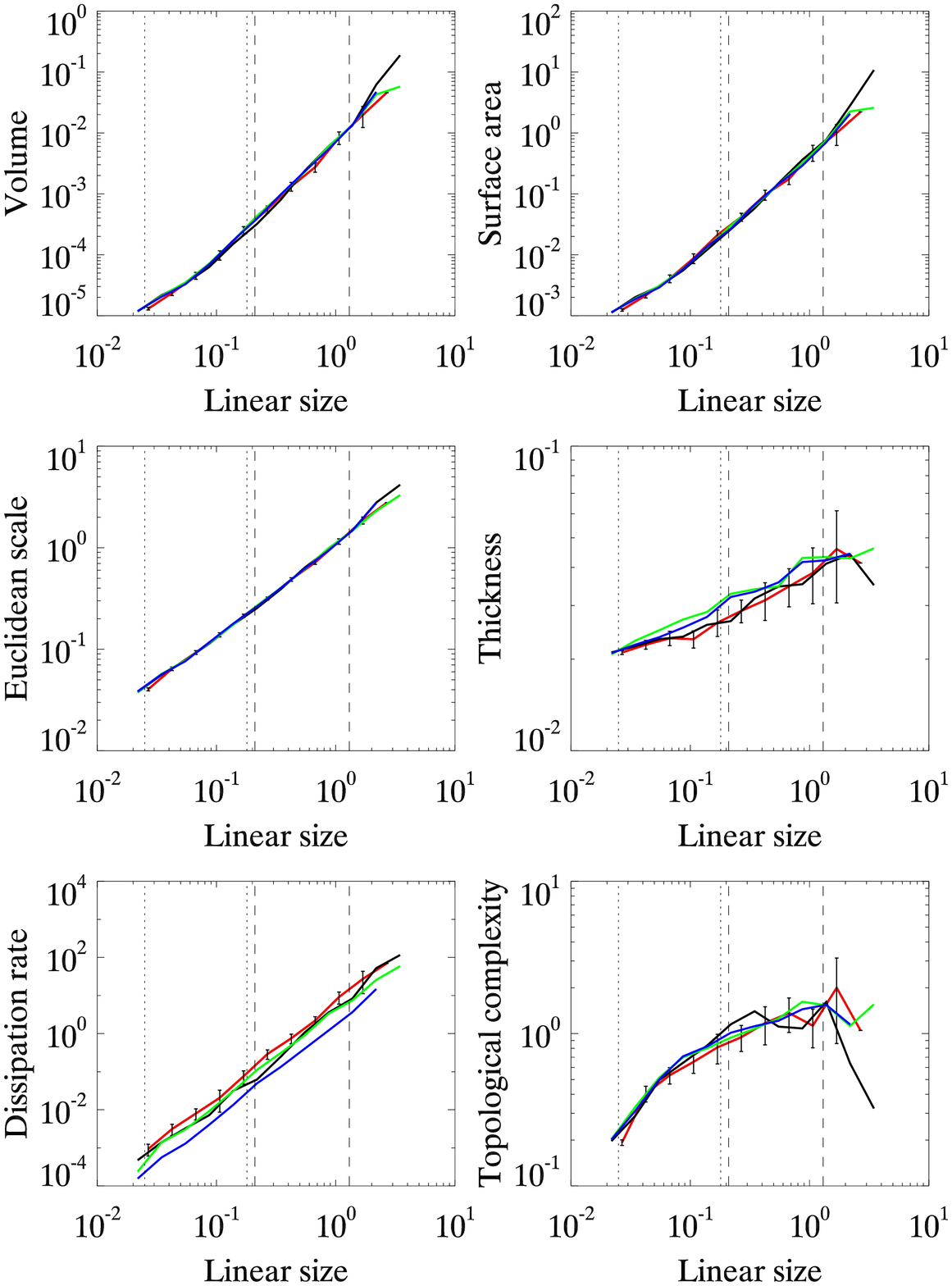}
\includegraphics[width=4.1cm, clip]{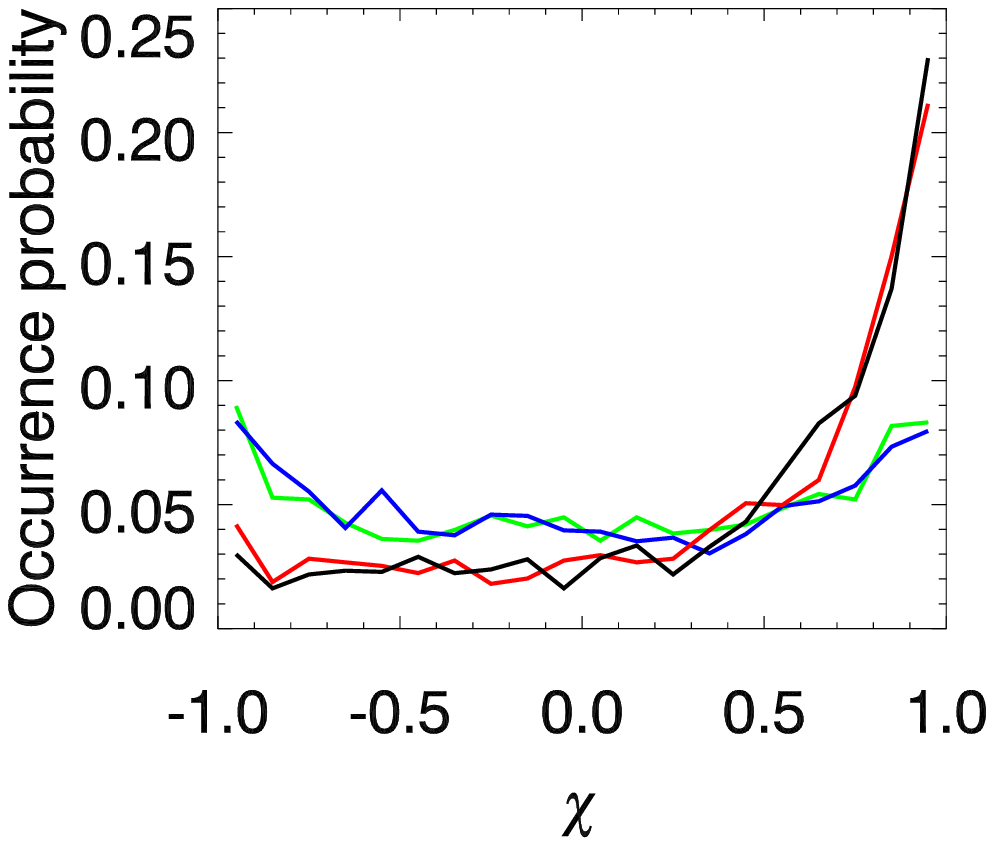}
\caption{({\it Color online})
Scaling characteristics of current and vorticity structures in ABC and OT flows (runs I and III, $N=512$, $t=4$, $a_{th}=2$) addressing the role of the velocity -- magnetic field alignment. In run I, structures in $j^2$ and $\omega^2$ are shown with light gray (green) and dark gray (blue) lines, whereas in run III,  they are given with middle gray (red) and black lines. Note a similarity in the volume and area statistics, and a significant difference in the alignment patterns of the two flows ($\chi$ is the cosine of the local alignment angle between ${\bf v}$ and ${\bf b}$). In addition to a stronger normalized $H_C$, the OT flow exhibits systematically thinner structures (smaller thickness for a given linear size, for both $j^2$ and $\omega^2$) compared to the ABC flow.} 
\label{f_vb_cos} \end{figure}

Figure \ref{high_Re_scaling} displays the numerical values of several key scaling exponents, reflecting the differences and the similarities between high- and low-Reynolds number ABC runs. The overlapping error bars indicate that the mean exponents are indistinguishable at the 99$\%$ confidence level. The plots confirm that the distribution exponents are roughly independent of $R_V$ whereas the geometric exponents tend to be larger in run II and are also closer to the value of two, as expected for idealized purely two-dimensional structures. The difference is consistent with the previously discussed observation that the current sheets in run II are considerably thinner and might therefore be better described by a 2D scaling model within the inertial range of scales.

Finally, note that the proximity of $D_R$ in run II to a value of unity indirectly indicates that the current sheets generated in this fluid are somewhat more isotropic in terms of their global (large-scale) spatial orientation compared to run I. For a fully isotropic macroscopic orientation, we expect the two linear measures, $R$ and $L$, to be directly proportional.

Overall, we conclude that the Reynolds number of the flow, at least in the case of unit magnetic Prandtl number ($\nu=\eta$), influences the geometry of the resulting dissipative structures, making them more 2D-like (higher aspect ratio) and better mixed in space, but it does not seem to alter their statistical properties as reflected by the family of the distribution functions we have studied. 

\subsection{The role of velocity-magnetic field correlations}

It has been known for quite some time that the amount of correlation between the velocity and the magnetic field plays a significant role in the dynamics of MHD turbulence (see  \cite{houches} for a review, and more recent works in \cite{servidio2, boldyrev}). This role can be global, altering the scaling of energy spectra when $H_C$, normalized by $E_T$, is strong, i.e., close to $\pm 1$; it can also be important, even when globally weak, since structures with strong alignment between ${\bf v}$ and ${\bf b}$ develop rapidly in a turbulent flow \cite{vb} (note that $H_C$ is not globally positive definite). It thus appears as a natural application of our detection algorithm to examine the properties of the selected clusters in this light.

Our examination of the data reveals a significant difference in the velocity - magnetic field alignment for the OT and ABC runs, in agreement with earlier analyses \cite{vb}. Defining $\chi$ as the cosine of the local alignment angle between ${\bf v}$ and ${\bf b}$, the OT run is characterized by an asymmetric $p(\chi)$ distribution (with a skewness of $\approx 1.0$), having a sharp maximum at $\chi= +1$. The alignment distribution suggests a prevailing parallel orientation of ${\bf v}$ and ${\bf b}$ fields inside the inertial range structures, with an average $\chi \approx 0.4$. The ABC run shows a nearly symmetric $p(\chi)$ distribution (skewness $\approx 0$) with the average alignment close to zero as well (see Fig. \ref{f_vb_cos}, bottom right panel). The stronger velocity -- magnetic field alignment in the OT fluid may be the primary reason for its distinct scaling behavior as represented in Tables  \ref{tab_inertial} -- \ref{tab_comp}. As we have already stressed, the OT $\tau$ exponents tend to be less sensitive to the crossover between the inertial and dissipative ranges of turbulence compared to the ABC run. This tendency is especially clear in the statistics of vorticity structures (black lines in Fig. \ref{f_vb_cos}, see also the last column in Table  \ref{tab_inertial}). At the same time, the geometric scaling of most of the parameters of $j^2$ and $\omega^2$ structures, in particular $V(L)$ and $A(L)$ dependencies, is practically the same for the two flows except for the geometric scaling of $H$, implying that the current and vorticity sheets are somewhat thinner in the OT run. 

Indeed, the alignment effect appears to have limited or no impact on scaling behavior of the intermittent structures. Surprisingly, the differences in the geometric and probabilistic scaling of OT and ABC turbulence are strikingly small compared to the dramatic difference in the $\chi$ distributions characterizing the two runs. It is also interesting that the $p(\chi)$ distributions have the same functional form for current and vorticity structures (for a given initial condition), suggesting the existence of a strong $j$ -- $\omega$ coupling correlated with the ${\bf v}$ - ${\bm b}$ alignment; this result may be linked to the fact that, when writing the MHD equations in terms of the Els\"asser fields ${\bf z}^{\pm}={\bf v}\pm {\bf b}$ for which the nonlinear terms reduce to an advection of one field by the other, one sees that the dynamics strongly couples the velocity and magnetic fields (and their derivatives) \cite{servidio2}. Other alignments could be considered \cite{servidio3}, from the point of view of structure analysis, basically those having a direct impact on the dynamics, such as the Lamb vector ${\bf v} \times \omega$, the Lorentz force ${\bf j} \times {\bf b}$ and Ohm's law ${\bf v} \times {\bf b}$. This is left for future work.

\section{A possible link with self-organized criticality}

The analysis presented in this section is motivated by the intensively debated connection between intermittent structures in turbulence and SOC, the latter having been discussed intensely  in the literature over the last decade. In particular, SOC has been proposed as the underlying physical mechanism responsible for the intermittency of the dissipation field in high-Reynolds number turbulent fluids \cite{sreen04,ur07,chang99,pacz05} (for a review of SOC in the context of Solar Wind and the magnetosphere, see, e.g., \cite{chapman2001, chang2003}). It has been suggested that dissipative regions can communicate over large distances, by analogy with critical avalanches in sandpile models of SOC, producing conditions for a statistically steady state of nonequilibrium critical behavior responsible for multifractal inhomogeneous dissipation \cite{chen03,chen04,bak05}.

To test the SOC avalanche hypothesis, one needs to obtain a collection of probability distribution functions describing the dissipative regions, and to study their scaling. The hallmark of SOC is the power-law shape of avalanche distributions over a number of parameters, some of which are studied here. According to the definition of SOC, avalanches are essentially spatiotemporal objects composed of all grid nodes involved in the formation of a given dissipative structure over its entire life cycle. Consequently, in order to rigorously verify SOC behavior in a turbulent fluid, one needs to detect and analyze its dissipative structures in both space and time. The results of our present analysis refer to static three-dimensional vorticity and current clusters observed at a fixed time; thus, these results cannot provide a definite answer to the question of whether or not incompressible MHD turbulence is related to SOC. Nevertheless, they can be used to make some preliminary estimates (see \cite{galtier} for an analysis of flaring activity in one space dimension, and \cite{einaudi} for the 2D case).
 
In the following derivation, we are assuming that the spatial intermittent structures explored in the previous sections are static snapshots of dynamic intermittent events evolving in space and time. Using the SOC approach, each of these events can be described by the spatiotemporal size $S$ representing the total number of grid nodes involved in the event over its life time $T$. The distributions of $S$ and $T$ are expected to scale as $p(S) \sim S^{-\tau_S}$ and $p(T)\sim  T^{-\tau_T}$. The avalanche size and lifetime scaling exponents ($\tau_S$ and $\tau_T$) are usually considered to be the primary measures of criticality defined by the universality class of a particular set of symmetries describing local interactions between the nodes \cite{benhur}.

Due to the absence of temporal dimension in our present analysis, neither of the two SOC exponents is directly accessible. However, by applying once again the conservation of probability, we can evaluate them indirectly through
\begin{eqnarray}
\tau_{S,\: T} &=& 1 + D_X(\tau_X-1)/D_{S, \:T}, 
%\tau_T &=& 1 + D_X(\tau_X-1)/D_T \ ,
\label{eq:tau} \end{eqnarray}
in which $S \sim L^{D_S}$, $T \sim L^{D_T}$, and $X$ stands for one of the static measures of the structures exhibiting power-law scaling as already discussed.

We start by analyzing in this light the dissipative subrange of scales. In the case of the ABC flow, the proximity of $\tau_X$ to the value 1 in that range (see Table \ref{tab_dissip}) makes uncertainty of $D_S$ and $D_T$ unimportant. For a wide range of $D_S$ and $D_T$ estimates presented in the SOC literature, and for various choices of $X$, Eq. (\ref{eq:tau}) predicts that the value of both the avalanche size and the lifetime distribution exponents for this flow are also close to unity. Thus, for instance, by plugging in $\tau_V$ and $D_V$ for current structures in the ABC flow \cite{note}, and using mean-field values for $D_S$ and $D_T$ (respectively 2 and 1) \cite{dhar}, one gets $\tau_S \approx \tau_T \approx  1$. Interestingly enough, the same calculation for the OT run yields a significantly different result, namely $\tau_S \approx 1.3$ and $\tau_T \approx 1.6$. 

\begin{table} 
\caption{Avalanche size and avalanche lifetime scaling exponents estimated in the dissipative range using the relations $\tau_{SX}=1 + D_X(\tau_X-1)/D_S$ and $\tau_{TX}=1 + D_X(\tau_X-1)/D_T$, with $X \in \{A, V, P\}$ (see Eq.~(\ref{eq:tau})) with the mean-field geometric exponents $D_S=2$ and $D_T=1$ and the $\tau_X$ exponents taken from Table III ($t=4$, $a_{th}=2$, $N=512$) }.
\begin{ruledtabular}
\begin{tabular}{lcccc}
          & Run I, $j^2$ &  Run I, $\omega^2$ & Run III, $j^2$ & Run III, $\omega^2$ \\
\hline
\\
%\multicolumn{5}{l}{\it Inertial range} \\
$\tau_{SA}$ & 0.97 & 1.03 & 1.29 & 1.19 \\ %
$\tau_{SV}$ & 1.08 & 1.12 & 1.31 & 1.28 \\ %
$\tau_{SP}$ & 0.98 & 1.15 & 1.43 & 1.36 \\ %
\\
$\left\langle \tau_{S} \right\rangle$ & 1.01 $\pm$ 0.06 & 1.10 $\pm$ 0.06 & 1.34 $\pm$ 0.08 & 1.28 $\pm$ 0.09 \\ %
%$\tau_L$  & 2.20 $\pm$ 0.22 & 2.57 $\pm$ 0.22 & 1.86 $\pm$ 0.23 & 1.94 $\pm$ 0.23 \\
\\
%\multicolumn{5}{l}{\it Dissipative range} \\
$\tau_{TA}$   & 0.93 & 1.06  &  1.58 &  1.37 \\ %
$\tau_{TV}$   & 1.15 &  1.25 &  1.62 & 1.56  \\ %
$\tau_{TP}$   & 0.96  &  1.30 &  1.86 &  1.73 \\ %
\\
$\left\langle \tau_{T} \right\rangle$  & 1.01 $\pm$ 0.12 & 1.20 $\pm$ 0.13 & 1.69 $\pm$ 0.15 & 1.55 $\pm$ 0.18 \\ %
%$\tau_L$  & 0.75 $\pm$ 0.12  & 0.82 $\pm$ 0.04  & 1.31 $\pm$ 0.21  &  1.16 $\pm$ 0.08 \\
\\
\end{tabular} \end{ruledtabular} \label{tab_soc}\end{table}

Table \ref{tab_soc} summarizes the estimated values of $\tau_S$ and $\tau_T$ exponents obtained using the relation (\ref{eq:tau}) in which we plug in the dissipative range scaling exponents characterizing volume, surface area, and energy dissipation rate in  $j^2$ and $\omega^2$ structures. We do not use the linear size exponents in this calculation since they are less reliable due to their dependence on the orientation of the structures (see the discussion of the small-scale anisotropy in section III.B). 

While the ABC avalanche exponents obtained from Eq. (\ref{eq:tau}) in the dissipative range are somewhat low compared to SOC exponents usually reported in the literature, the OT exponents clearly fall within the range of values expected for many SOC sandpiles. Thus, for example, they are a very close match to the 2D realization of the directed Abelian sandpile model (DASM) \cite{dhar}, an exactly solvable version of the paradigmatic Bak-Tang-Wiesenfeld model \cite{BTW}. The DASM avalanche distributions are described by the exponents $\tau_s = 4/3$ and $\tau_T = 3/2$ which are nearly the same as the OT values reported in Table V. Several other avalanching models are approximately consistent with the predicted OT exponents, such as, e.g., the Manna two-state model \cite{manna}, the Bak-Snappen model of punctuated evolution \cite{bak}, and the absorbing state phase transition model \cite{vesp}, all in two spatial dimensions. The distinctive feature of DASM which might be responsible for the best match with critical behavior in the dissipative range of the OT flow is the existence of the preferred spatial direction in which the avalanche fronts propagate. As we have mentioned earlier in the text, the local spatial anisotropy seems to be a significant factor in the dissipative range scaling of the studied fluids, and therefore the observed agreement between OT and DASM avalanche exponents may be more than just a coincidence.
 
\begin{figure}
\includegraphics[width=8.5cm, clip]{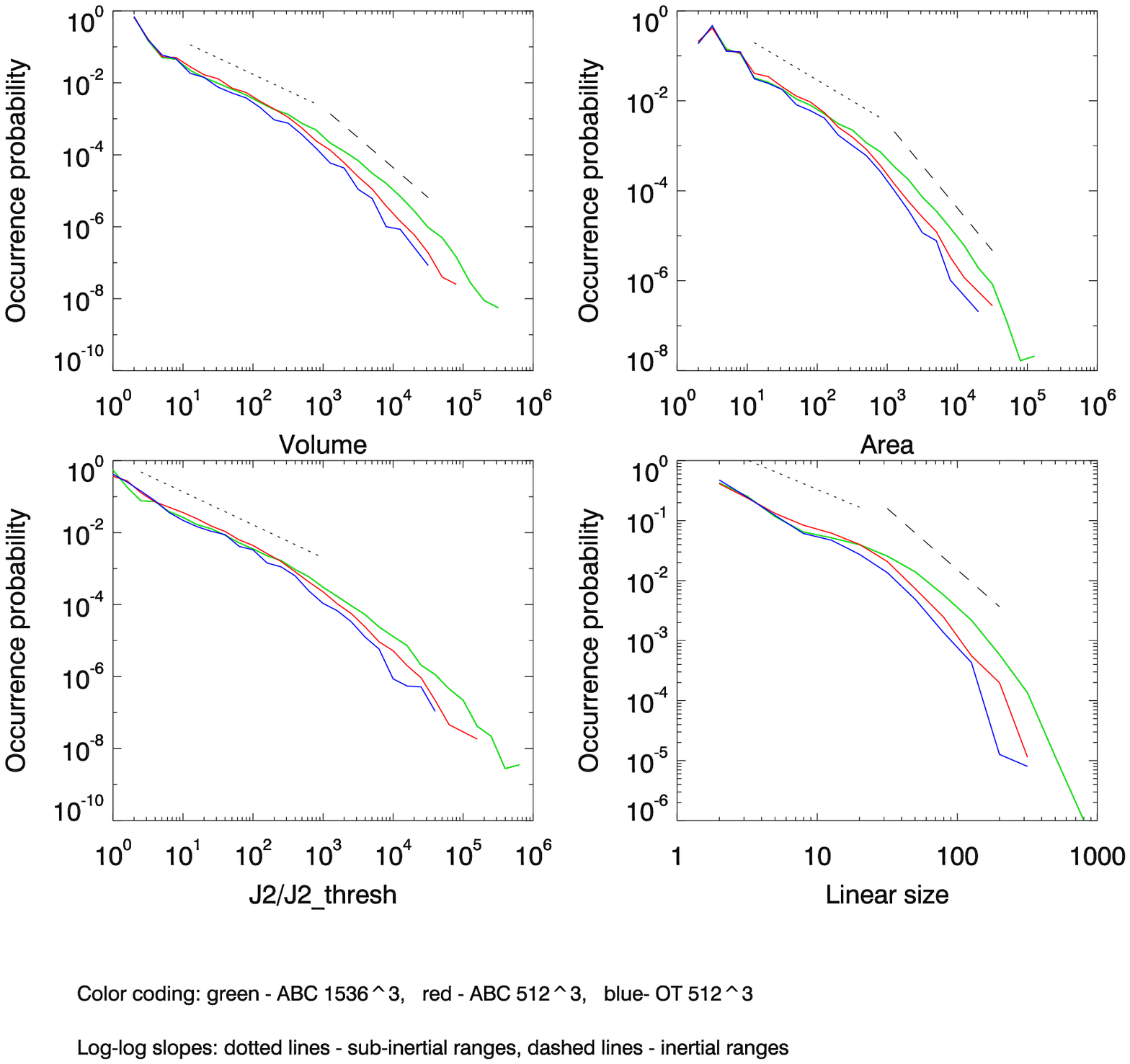}
\caption{({\it Color online})
Probability distributions of the dimensionless volume and area in runs I (middle gray or red), II (light gray or green), and III (dark gray or blue), demonstrating a wider range of small-scale power-law behavior in the high $R_V$ (high-resolution) simulation.}
\label{f_soc} \end{figure}

Besides the existence of well-known SOC universality classes consistent with the exponents obtained from the analysis, a significant piece of evidence for the involvement of SOC dynamics in the formation of multifractal intermittent dissipation fields comes from the statistics of dimensionless measures of dissipative structures representing their size and geometry in terms of discrete grid nodes, by analogy with sandpile simulations. A predicted effect for a SOC system is an expansion of the range of the power-law scaling of such measures for increasingly large lattice sizes, known as the critical finite-size scaling (FSS) behavior.

Our analysis shows that the probability distributions of dimensionless quantities of turbulent structures (e.g., dimensionless volume $V/\delta^3$ and area $A/\delta^2$; see Fig.~\ref{f_soc}) exhibit signatures of FSS, assuming that SOC avalanches do develop in the dissipative (sub-inertial) range of scales. A comparison of dimensionless distributions in high- and low-resolution runs shows that the range of power-law scaling describing the smallest structures, and thus involving a limited number of grid nodes, expands towards larger scales as $N$ increases from 512 to 1536. This type of behavior is among the most distinctive signatures of SOC systems. It indicates that the intrinsic mechanisms of avalanching dynamics are scale-free with no limit other than the limited size of the Reynolds number. In the theory of critical phenomena, the tendency seen in Fig. \ref{f_soc} is usually approximated by truncated power-law distributions of the form
\begin{equation}
p(X) = X^{-\tau_X}\,f(X/X_c) \ ,
\end{equation}
where $f$ is an appropriate scaling function and $X_c$ the (apparent) characteristic scale of structures increasing with a discrete system size $N$ as $X_c \sim N^{\kappa_X}$. Due to the scale-free nature of SOC states, we also expect that $\kappa_X$ is close to the corresponding geometric exponent $D_X$ (because the largest linear scale associated with $N$ has the same effect on the distributions of structure sizes as the intermediate scales associated with $L$). However, it should be recognized that unlike FSS in ordinary SOC simulations with simple boundary conditions, the upper scale of the presumed SOC behavior in our turbulent runs is controlled by a complex process -- the inertial range turbulent cascade. Consequently, the functional form of the scaling function $f$ in a turbulent fluid should perhaps include a scale-free component accounting for the fluid turbulence at larger scales (in contrast, e.g., to the exponentially decaying $f$ commonly used in sandpile simulations).

The interpretation of SOC exponents found in the inertial range is more ambiguous. Technically, they are not far from the values $\tau_S=1.5$ and $\tau_T=2$ describing SOC dynamics in 3-dimensional stochastic and deterministic directed sandpiles (see \cite{hughes} and refs. therein). However, this similarity can be misleading as the above-mentioned models possess a distinct geometry consisting of two spatial dimensions in which dissipative events can grow isotropically and one dimension allowing for unidirectional (directed) growth only. Whether or not the growth of dissipative structures in the inertial-range MHD turbulence contains such a preferred direction implying strong mesoscopic anisotropy, remains to be verified. Until then, we only associate the dissipative range with SOC behavior. 
%do not match any SOC class of universality we know of, and therefore for the moment we only associate the dissipative range with SOC behavior. 
It may be the case that the accuracy of the present analysis in the inertial range is insufficient, or that SOC behavior is only limited to the dissipative range. A direct spatio-temporal analysis or growing and decaying dissipative structures will be instrumental for validating our SOC observations in the inertial range and for reducing the uncertainties in the exponents. Such analysis will also allow for a study of whether the ergodicity assumptions often used in the study of turbulent flows are valid \cite{matthaeus_1f}.

Overall, the results obtained in this paper suggest that small-scale dissipative structures, observed below the smallest inertial range scale, are associated with SOC. If this hypothesis is correct, the small-scale intermittency in 3D MHD turbulence can be interpreted as a propagation of local instabilities from small to large scales indicative of SOC avalanches. This propagation should reflect a tendency of the smallest dissipative structures, such as, e.g., current filaments, to merge into larger clusters in an avalanche fashion, and it does not necessarily imply a transport of energy in Fourier space in the opposite direction as to the main (direct) turbulent cascade of energy. One could think for example of larger dissipation events, like major flares in the solar corona, emerging from a cooperative behavior of smaller events (e.g., nanoflares), an effect reminiscent of non-locality of nonlinear interactions between widely separated scales \cite{nonlocal}. This possibility has been discussed intensively in the literature, in particular in the framework of forest-fire models of multifractal inhomogeneous dissipation in turbulent media \cite{chen}. Our results seem to be the first (but so far indirect) evidence of such behavior in incompressible MHD observed through DNS in 3D, even though, of course, SOC behavior in MHD has been advocated for a long time following the pioneering paper of Lu and Hamilton \cite{lu} (see also \cite{einaudi,chang99,chapman07,ur07}).

In a more general context, to fully describe SOC behavior in turbulence one would have to understand several fundamental aspects shaping the dynamics of the intermittent structures. For example, what plays the role of a threshold in MHD turbulence, so that avalanches (dissipative events) of various sizes can happen? It may be a collective effect triggered by sweeping of large scales, pushing together magnetic field lines of opposite polarities to come in close contact as has been proposed by Klimas et al. \cite{klimas}, or it may be the instability of current (or vorticity) sheets below a given thickness (at a fixed viscosity/resistivity). If SOC is indeed identified, other relevant questions involve, e.g., what is the underlying phase transition, what is the order parameter, and what field can be associated with the susceptibility that diverges at the critical point? These questions may not have clear answers, as the identification of criticality in fluid and MHD turbulence is not straightforward, although an example in the context of atmospheric precipitation in tropical convection has been put forward recently \cite{peters}.

We can speculate on one of the possible sources of burstiness in the dissipation and its relation with avalanches. Starting from the pioneering work of Onsager, Lee and Kraichnan \cite{onsager,tdlee,rhk}, one can think of the dynamical evolution of a turbulent flow being due to nonlinear interactions with weak forcing and weak dissipation balancing each other. Solutions of the ideal truncated equations obtain at late times in the simplest instance equipartition between all the modes, with zero energy flux. At intermediate times and intermediate scales, one observes turbulent dynamics with non-zero flux \cite{euler_meb}, the ``dissipation'' of large-scale energy being associated to a turbulent eddy viscosity due to the thermalized modes at small scale (see \cite{euler_hel,euler_mhd2d} for similar results for helical flows, and for 2D MHD). This dynamics have also been observed in viscous cases, e.g., in  Navier-Stokes at high resolution, where the resulting flow can be decomposed into a set of coherent structures with a spectrum close to Kolmogorov, and a large number of modes at small scales in thermal equilibrium \cite{okamoto}.

Related with these results, it has been known for some time, and in different instantiations of turbulent flows, that the energy flux of a given sign on average, has in fact huge fluctuations of both signs and of amplitudes much larger than the mean (of order unity for characteristic velocities and lengths $U_0=1$ and $L_0=1$, respectively; see for example \cite{connaugthon} and references therein, and \cite{graham} for studies of regions with zero flux in models of turbulence). These large fluctuations in the flux can be attributed to the balance between forcing and dissipation mentioned above, and to the two components (one thermalized and random, one turbulent and coherent) identified in turbulent flows at small scales. The interplay between the two components can result in a bursty flux transfer of energy to the small scales, as observed in particular when looking at dissipation and reconnection events \cite{politano89}. These bursts are the needed excursions that lead the system away from equilibrium and may give rise to a state of criticality, in order to dissipate the energy accumulated over various lapses of time through the injection mechanism. Some of these events will trigger other events, by pushing around structures that through their associated pressure fields can make contact with other structures that may in turn destabilize.

Finally, it is worth mentioning that intermittent bursts of energy transfer have been observed  in MHD in the so-called shell models for turbulent flows  \cite{gloaguen}. These models can be viewed as a poor-man template, set on a lattice, for the temporal evolution of the Navier-Stokes or MHD equations (see \cite{carbone} and references therein for a recent review); in the simplest case, one retains only nearest-neighbor interactions which are built in such a way that the quadratic invariants are preserved. Shell models have been known to exhibit avalanche behavior as well (see \cite{boffetta99, carbone} for a discussion), although SOC interpretation of energy avalanches in such simplified models of turbulent cascade remains questionable, as there are indications that only spatially-distributed systems with clear time-scale separation between the driving and dissipation mechanisms are able to exhibit robust SOC signatures \cite{uritsky09}.

\section{Conclusion}

\subsection{Summary of the results}

In this paper, we have analyzed three sets of data stemming from high-Reynolds number numerical simulations of MHD turbulence in three dimensions at a magnetic Prandtl number of unity; periodic boundary conditions are assumed and there is no imposed uniform magnetic field nor is there any forcing. Initial conditions are either the so-called 3D Orszag-Tang vortex (OT) or the Arn'old-Beltrami-Childress (ABC) flow; the two flows have different velocity-magnetic field correlation $H_C$. Numerical resolutions range from $512^3$ grid points to $1536^3$, with Taylor Reynolds numbers $R_{\lambda}$ varying from $630$ to $1100$.

We find that current and vorticity sheets behave in similar fashion and that, overall, the probabilistic properties of these structures do not depend on either $R_{\lambda}$ or the normalized value of $H_C$, i.e., the degree to which ${\bf v}$ and ${\bf b}$ are globally aligned. Other factors like the detection threshold used for the analysis (within 1 to 3 standard deviations above the mean) or the time at which the turbulent fluid is analyzed (after reaching the peak of the dissipation) appear to be  irrelevant as well.

As expected, dissipative structures are thinner and more complex (and closer to being two-dimensional) the higher the Reynolds number. Compared to the ABC flow, the OT run has more efficient kinetic energy dissipation (higher dissipation rate per unit volume) and its distributions of structure parameters are closer to a ``monofractal'' shape across all scales, i.e., approximately described by a single power-law. The high $R_{\lambda}$ run has a higher dissipation rate per unit volume (but roughly the same rate per unit surface), and essentially the same form of probability distributions as in the low $R_{\lambda}$ run.

The inertial scaling exponents characterizing the OT and ABC flows are similar, with differences between the exponents more marked at smaller scales. The exponents obtained at these scales (in the dissipative subrange) can be associated with SOC universality classes. Our findings suggest that whereas the inertial-range scaling is more likely to be dictated by MHD turbulent cascades (energy spectra, and structure functions in general), the dissipative range of scales may be governed by self-organized criticality; this is perhaps the main finding that stems from our cluster analysis.

Why is the SOC scaling found in the dissipation range? Simply because this is where the approximately ideal dynamics breaks down; at the small-scale end of the inertial range detailed conservation of quadratic invariants is violated by non-zero dissipation. In fact, if there are singularities in such flows, the dissipation can be order unity (in terms of the characteristic velocity and length). Such exchanges must be bursty insofar as they are concentrated on a small scale; they are rare as they only occur in special cases, and thus they must be strong as they provide the dissipation on average to balance the energy injected by the larger scales. In other words, the stochastic (and irreversible) element comes from the fact that dissipative events occur where vortices and currents of opposite polarities meet, at random time because of the randomness of large-scale structures for times longer than the eddy-turnover time (see, e.g., \cite{matthaeus_1f} for studies of $1/f$ noise in turbulent flows). This results in strong bursts when the system undergoes dissipation events (such as reconnection) with reconfiguration of the fields. In this simplified picture, the critical parameter for transition may be the Kolmogorov dissipation length $\ell_{diss}$, in scale-space, or rather the dimensionless local Reynolds number $u_{\ell} \ell/\nu$ of order unity, by definition, at $\ell\sim \ell_{diss}$.

\subsection{Final remarks}

In conclusion, we have shown that the cluster algorithm presented in section II.B and used throughout this study can easily and systematically detect a multitude of current and vorticity structures in a turbulent flow,  with a large dynamical range in their intensity, and that it readily leads to an analysis of their relevant physical properties. As important perhaps, one can then study some other statistics of such structures, such as the alignment between velocity and magnetic field. Although the analysis presented in this paper deals with properties of decaying flows in the vicinity of the peak of dissipation, one can note that such flows are thought to behave similarly to the statistically steady forced case, since their dynamics is quasi-steady for some interval of time around that peak (see, e.g., \cite{1536b}). It would however be of interest to study the statistically stationary case of MHD turbulence; this is left for future work.

Relating the scaling exponents found in a given analysis to other, more traditional measures of complexity in a turbulent flow dealing, e.g., with correlation functions, is not necessarily a straightforward task. Some fascinating results concerning the behavior of two-dimensional Navier-Stokes turbulence have been unraveled recently \cite{falko} with no clear direct connection to the scaling of, say, the energy spectrum; in this 2D case with an inverse cascade of energy to large scales, the study of the zero-line vorticity contours led to the discovery of  a link with a specific class of percolation and anomalous diffusion through the scaling laws for, e.g., length versus diameter. The fact that the dynamics of turbulent flows contains elements of critical phenomena and conformal invariance associated with invariance properties and symmetry groups of the underlying equations points to the need to further our studies of such flows using scaling tools.

There are other algorithms that can examine coherent structures in turbulent flows, following the pioneering work for 2D Navier-Stokes fluids \cite{mcwilliams}. Prominent among them nowadays is the wavelet decomposition \cite{farge} which leads to a vision of turbulent flows in three dimensions as a set of coherent structures (vortex filaments in the fluid case) with a Kolmogorov spectrum, together with incoherent quasi-Gaussian eddies which contain most of the degrees of freedom and which are in some sense slaved to the coherent structures; a similar analysis has been done recently in MHD \cite{okamoto}. The use of visualization techniques and lossless compression of data is also a possible tool to analyze turbulent structures \cite{vapor,mprast}. Combining such tools with the analysis of hypercubes of data taking into account the temporal dimension of structures may prove fruitful, but in three dimensions this represents a challenge that we want to tackle in the near future, both for fluids and MHD; it will allow for a better connection between turbulence, intermittency, structures and self-organized criticality.

\acknowledgments
NCAR is sponsored by NSF. PDM acknowledges support from grants UBACYT X468/08 and PICT 2007-02211. The work of VMU and ED was supported by the Canadian Space Agency. VMU thanks G. Uritsky for his technical help in preparing this manuscript.


\begin{thebibliography}{99}
\bibitem{note}
Note that the volume of the structures in our definition is an instantaneous value of the avalanche size $S$ at a given time, and is therefore a natural choice in frames of this calculation.
\end{thebibliography}
\end{document}